\documentclass{article}
\input{style.sty}
\usepackage[utf8]{inputenc}

\title{\textbf{5G System Security Analysis }}
\author{Gerrit Holtrup$^{1}$; William Lacube$^{2}$; Dimitri Percia David$^{3}$; Alain Mermoud$^{2}$; \\\ G\'{e}r\^{o}me Bovet$^{2}$; Vincent Lenders$^{2}$\\}
\date{}

\begin{document}

\maketitle
\thispagestyle{empty}

\begin{center}
    Cyber-Defence Campus
\end{center} 

\vspace{0.75cm}

\noindent
\scriptsize $^{1}$ Kudelski IoT Security, Switzerland\\
\noindent
\scriptsize $^{2}$ armasuisse, Science and Technology, Switzerland\\
\noindent
\scriptsize $^{3}$ Information Science Institute, Faculty of Economics and Management, University of Geneva, Switzerland\\

\vspace{1cm}

\section*{Abstract}
\normalsize	
\noindent
\textit{Fifth generation mobile networks (5G) are currently being deployed by mobile operators around the globe. 5G acts as an enabler for various use cases and also improves the security and privacy over 4G and previous network generations. However, as recent security research has revealed, the standard still has security weaknesses that may be exploitable by attackers. In addition, the migration from 4G to 5G systems is taking place by first deploying 5G solutions in a non-standalone (NSA\label{NSA}) manner where the first step of the 5G deployment is restricted to the new radio aspects of 5G, while the control of the user equipment is still based on 4G protocols, i.e.  the core network is still the legacy 4G evolved packet core (EPC\label{EPC}) network. As a result, many security vulnerabilities of 4G networks are still present in current 5G deployments. This paper presents a systematic risk analysis of standalone and non-standalone 5G networks. We first describe an overview of the 5G system specification and the new security features of 5G compared to 4G. Then, we define possible threats according to the STRIDE threat classification model and derive a risk matrix based on the likelihood and impact of 12 threat scenarios that affect the radio access and the network core. Finally, we discuss possible mitigations and security controls. Our analysis is generic and does not account for the specifics of particular 5G network vendors or operators. Further work is required to understand the security vulnerabilities and risks of specific 5G implementations and deployments.}  \\

\vspace*{1.5cm}

\begin{center}
    19 August 2021
\end{center} 

\vspace*{\fill}
\noindent
\textbf{Corresponding author:} alain.mermoud@ar.admin.ch \\ \\
\noindent
\textbf{Disclaimer:} Opinions expressed in this report are solely our own and do not express the views or opinions of our employer or the government.

\clearpage
\pagenumbering{arabic}
\setcounter{page}{2}

\setcounter{tocdepth}{2}
\tableofcontents

\clearpage
\listoffigures
\listoftables

\clearpage
\section{List of Abbreviations}

{\scriptsize
\begin{xltabular}{\textwidth}{lXXm{1.5cm}}
Acronym & Stands For                                  & Definition                                                                                                                                                                                                           & First occurrence page \\ \hline
3GPP    & $3^{rd}$ Generation Partnership Project     &                                                                                                                                                                                                                      & \pageref{3GPP}        \\
5GC     & 5G Core network                             & Control plane implementation following the 5G specifications.                                                                                                                                                        & \pageref{5GC}        \\
5G-NR   & 5G New Radio                                &                                                                                                                                                                                                                      & \pageref{5G-NR}        \\
5G-GUTI & 5G Globally Unique Temporary Identifier  & Globally unique identifier allocated to a UE after an initial registration procedure.                                                                                                                               & \pageref{5G-GUTI}        \\
ABBA    & Anti-Bidding down Between Architectures     &                                                                                                                                                                                                                      & \pageref{ABBA}        \\
AES     & Advanced Encryption Standard                & Symmetric-key algorithm.                                                                                                                                                                                              & \pageref{AES}        \\
AEAD    & Authenticated Encryption with Associated Data &                                                                                                                                                                                                                    & \pageref{AEAD}        \\
AS      & Access Stratum                              &                                                                                                                                                                                                                      & \pageref{AS}        \\
AMF     & Access and Mobility Management Function     & Definition on page \pageref{sec4.1.3}.                                                                                                                                                                                & \pageref{AMF}        \\
AUSF    & Authentication Server Function              & Definition on page \pageref{sec4.1.4}.                                                                                                                                                                                & \pageref{AUSF}        \\
BFN     & Beam Forming Networks                       &                                                                                                                                                                                                                      & \pageref{BFN}        \\
CU      & Central Unit                                &                                                                                                                                                                                                                      & \pageref{CU}        \\
D2D     & Device To Device                            &                                                                                                                                                                                                                      & \pageref{D2D}        \\
DHE     & Diffie–Hellman key Exchange                 & A method for securely exchanging cryptographic keys over a public channel.                                                                                                                                           & \pageref{DHE}        \\
DoS     & Denial Of Service                           &                                                                                                                                                                                                                      & \pageref{DoS}        \\
DU      & Distributed Unit                            &                                                                                                                                                                                                                      & \pageref{DU}        \\
ECDH    & Elliptic-curve Diffie–Hellman               & A key agreement protocol that allows two parties, each having an elliptic-curve public–private key pair, to establish a shared secret over an insecure channel.                                                      & \pageref{ECDH}        \\
ECIES   & Elliptic Curve Integrated Encryption Scheme & Incarnation of the Integrated Encryption Scheme (IES).                                                                                                                                                               & \pageref{ECIES}        \\
ECU     & Electronic Control Unit                    & Electronics component inside a vehicle architecture.                                                                                                                                                                 & \pageref{ECU}        \\
eMBB    & Enhanced Mobile BroadBand                  & Standardized slice/service type for "normal"  communications.                                                                                                                                                        & \pageref{eMBB}        \\
EN-DC   & E-UTRA-NR Dual Connectivity                &                                                                                                                                                                                                                      & \pageref{EN-DC}        \\
en-gNB  &                                             & gNB in a NSA context.                                                                                                                                                                                                & \pageref{en-gNB}        \\
eNB     & Evolved NodeB                               &                                                                                                                                                                                                                      & \pageref{eNB}        \\
ECC     & Elliptic-curve cryptography                 & An approach to public-key cryptography based on the algebraic structure of elliptic curves over finite fields.                                                                                                       & \pageref{ECC}        \\
EPC     & Evolved Packet Core                        & 4G core network.                                                                                                                                                                                                     & \pageref{EPC}        \\
EPS     & Evolved Packet System                       &                                                                                                                                                                                                                      & \pageref{EPS}        \\
ESP     & Encapsulating Security Payload              & Member of the IPsec protocol suite.                                                                                                                                                                                   & \pageref{ESP}        \\
FIB     & Focused Ion Beam                            & Scientific instrument.                                                                                                                                                                                                & \pageref{FIB}        \\
gNB     & NR Node B                                   & 5G RAN implementation.                                                                                                                                                                                               & \pageref{gNB}        \\
HSM     & Hardware Security Module                    & Physical computing device that safeguards and manages digital keys, performs encryption and decryption functions.                                                                                                     & \pageref{HSM}        \\
HTTP    & Hypertext Transfer Protocol                 & Application layer protocol for distributed, collaborative, hypermedia information systems.                                                                                                                            & \pageref{HTTP}        \\
HW      & Hardware                                    &                                                                                                                                                                                                                      & \pageref{HW}           \\
IKEv2   & Internet Key Exchange version 2             & Protocol used to set up a security association (SA) in the IPsec protocol suite.                                                                                                                                     & \pageref{IKEv2}        \\
IMEI    & International Mobile Equipment Identity  & Unique identifier of the mobile equipment that is part of the UE together with the USIM.                                                                                                                            & \pageref{IMEI}        \\
IMSI    & International Mobile Subscriber Identity &                                                                                                                                                                                                                      & \pageref{IMSI}        \\
IoT     & Internet of Things                         &                                                                                                                                                                                                                      & \pageref{IoT}        \\
IP      & Internet Protocol                           & Principal communications protocol in the Internet protocol suite for relaying datagrams across network boundaries.                                                                                                    & \pageref{IP}        \\
IPsec   & Internet Protocol Security                  & Secure network protocol suite that authenticates and encrypts the packets of data to provide secure encrypted communication between two computers over an Internet Protocol network.                                 & \pageref{IPsec}        \\
IPX     & Internetwork Packet Exchange                & Network layer protocol in the IPX/SPX protocol suite.                                                                                                                                                                 & \pageref{IPX}        \\
Ka band & Kurz-Above band                             & Frequencies in the range 26.5–40 gigahertz (GHz).                                                                                                                                                                     & \pageref{Ka band}        \\
LTE     & Long Term Evolution                         & Standard for wireless broadband communication for mobile devices and data terminals.                                                                                                                                 & \pageref{LTE}        \\
M2M     & Machine To Machine                         &                                                                                                                                                                                                                      & \pageref{M2M}        \\
MCU     & MicroController Unit                        & Small computer on a single chip.                                                                                                                                                                                     & \pageref{MCU}        \\
ME      & Mobile Equipment                           & Part of the UE identified through the PEI.                                                                                                                                                                            & \pageref{ME}        \\
MEC     & Mobile Edge Cloud                           & For example a data endpoint close to the radio network to reduce latency in the communications pipe.                                                                                                                 & \pageref{MEC}        \\
MIMO    & Multiple Input Multiple Output              &                                                                                                                                                                                                                      & \pageref{MIMO}        \\
MIoT    & Massive IoT                                 & Standardized slice/service type for mMTC devices (cf. 5.15.2.2 in \cite{ref7}).                                                                                                                                          & \pageref{MIoT}        \\
MME     & Mobility Management Entity                  & Equipment that manages the control plane in a 4G network.                                                                                                                                                             & \pageref{MME}        \\
mMTC    & Massive Machine Type Communications        &                                                                                                                                                                                                                      & \pageref{mMTC}        \\
MR-DC   & MultiRadio Dual Connectivity               &                                                                                                                                                                                                                      & \pageref{MR-DC}        \\
N3IWF   & Non-3GPP InterWorking Function            & Function connecting a non-3GPP access network to the 5GC.                                                                                                                                                            & \pageref{N3IWF}        \\
NAS     & Non-Access Stratum                         &                                                                                                                                                                                                                      & \pageref{NAS}        \\
NDS     & Network Domain Security                     & NDS/IP utilizes IP Security (IPSec) to implement security domain services.                                                                                                                                           & \pageref{NDS}        \\
NEA     & 5G Encryption Algorithm                     &                                                                                                                                                                                                                      & \pageref{NEA}        \\
NEF     & Network Exposure Function                 & 5G network function exposing limited features to third party functions outside the 5GC, Definition on page \pageref{sec4.1.9}.                                                                                       & \pageref{NEF}        \\
NIA     & 5G Integrity Algorithm                      &                                                                                                                                                                                                                      & \pageref{NIA}        \\
NG-RAN  & Next-Generation RAN                         &                                                                                                                                                                                                                      & \pageref{NG-RAN}        \\
NGAP    & Next Generation Application Protocol      & Protocol used on N2 interface between the UE/gNB and the AMF.                                                                                                                                                        & \pageref{NGAP}        \\
NF      & Network Function                           &                                                                                                                                                                                                                      & \pageref{NF}        \\
NRF     & Network Repository Function               & Definition on page \pageref{sec4.1.9}.                                                                                                                                                                                & \pageref{NRF}        \\
NSA     & Non Stand Alone                             & Hybrid deployment of 5G gNBs interfacing with a 4G core network.                                                                                                                                                     & \pageref{NSA}        \\
NSSF    & Network Slice Selection Function          & Definition on page \pageref{sec4.1.9}.                                                                                                                                                                                & \pageref{NSSF}        \\
PEI     & Permanent Equipment Identifier            & 5G term for the equipment identifier that in the normal case is equal to the IMEI or IMEISV.                                                                                                                         & \pageref{PEI}        \\
PCF     & Policy Control Function                     & Definition on page \pageref{sec4.1.9}.                                                                                                                                                                                & \pageref{PCF}        \\
PDU     & Protocol Data Unit                          & Logical connection.                                                                                                                                                                                                   & \pageref{PDU}        \\
PLMN    & Public Land Mobile Network                  & Combination of wireless communication services offered by a specific operator in a specific country.                                                                                                                  & \pageref{PLMN}        \\
QoS     & Quality of Service                          & Description or measurement of the overall performance of a service.                                                                                                                                                   & \pageref{QoS}        \\
RAM     & Random Access Memory                        & Volatile type of memory.                                                                                                                                                                                              & \pageref{RAM}        \\
RAN     & Radio Access Networks                       &                                                                                                                                                                                                                      & \pageref{RAN}        \\
RSA     & Rivest–Shamir–Adleman                       & Public-key cryptosystem.                                                                                                                                                                                              & \pageref{RSA}        \\
RRC     & Radio Resource Control                      & Layer 3 (Network Layer) protocol.                                                                                                                                                                                     & \pageref{RRC}        \\
SA      & Standalone Access                           &                                                                                                                                                                                                                      & \pageref{SA}        \\
SCA     & Small Cell Access                           &                                                                                                                                                                                                                      & \pageref{SCA}        \\
SDR     & Software Defined Radio                      &                                                                                                                                                                                                                      & \pageref{SDR}        \\
SEAF    & SEcurity Anchor Function                   & 5G function located inside the AMF.                                                                                                                                                                                   & \pageref{SEAF}        \\
SEM     & Scanning Electron Microscope                & Scientific instrument.                                                                                                                                                                                                & \pageref{SEM}        \\
SEPP    & Security Edge Protection Proxy             & Definition on page \pageref{sec4.1.5}.                                                                                                                                                                                & \pageref{SEPP}        \\
SIM     & Subscriber Identification Module            &                                                                                                                                                                                                                      & \pageref{SIM}        \\
SMF     & Session Management Function                 & Definition on page \pageref{sec4.1.7}.                                                                                                                                                                                & \pageref{SMF}        \\
SQN     & Sequence Number                             &                                                                                                                                                                                                                      & \pageref{SQN}        \\
SUPI    & Subscription Permanent Identifier         & Generic name of the IMSI in 5G networks as unique identifier of a user’s identity both for 3GPP and non-3GPP equipment.                                                                                             & \pageref{SUPI}        \\
SUCI    & Subscription Concealed Identifier         & Version of the SUPI that is protected using asymmetric cryptography. The keys used for the concealment are issued by the home network of the user and the public key part is stored inside the UICC of the device. & \pageref{SUCI}        \\
SW      & Software                                    &                                                                                                                                                                                                                      & \pageref{SW}        \\
TCP     & Transmission Control Protocol               & Main protocols of the Internet protocol suite.                                                                                                                                                                       & \pageref{TCP}        \\
TLS     & Transport Layer Security                   & Standard Internet security protocol for the authentication and encryption of data.                                                                                                                                   & \pageref{TLS}        \\
UDM     & Unified Data Management                    & Definition on page \pageref{sec4.1.6}.                                                                                                                                                                                & \pageref{UDM}        \\
UDP     & User Datagram Protocol                      &                                                                                                                                                                                                                      & \pageref{UDP}        \\
UDR     & Unified Data Repository                    & Definition on page \pageref{sec4.1.9}.                                                                                                                                                                               & \pageref{UDR}        \\
UE      & User Equipment                              & Mobile terminal comprised of the USIM and the mobile equipment.                                                                                                                                                      & \pageref{UE}        \\
UICC    & Universal Integrated Circuit Card         & Physically secure device that will host the USIM Inside the UE.                                                                                                                                                      & \pageref{UICC}        \\
UPF     & User Plane Function                        & Definition on page \pageref{sec4.1.8}.                                                                                                                                                                                & \pageref{UPF}        \\
URLLC   & Ultra Reliable Low Latency Communications  & Standardized slice/service type for devices requiring guaranteed QoS with low packet loss and low latency.                                                                                                           & \pageref{URLLC}        \\
USIM    & Universal Subscriber Identity Module      & Module storing the identity (SUPI) and associated keys of a user inside the mobile equipment.                                                                                                                        & \pageref{USIM}        \\
V2X     & Vehicle To Everything                      &                                                                                                                                                                                                                      & \pageref{V2X}        \\
V2V     & Vehicle To Vehicle                          &                                                                                                                                                                                                                      & \pageref{V2V}        \\
VM      & Virtual Machine                             & Virtualization/emulation of a computer system.                                                                                                                                                                        & \pageref{VM}         \\
VNF     & Virtualized Network Function              & Network function implemented without using dedicated hardware and potentially being implemented as a cloud instance.                                                                                                & \pageref{VNF}        \\
ZUC     &                                             & Stream cipher.                                                                                                                                                                                                        & \pageref{ZUC}        \\
\end{xltabular}
}

\clearpage
\section{Introduction}
\label{intro}
The arrival of fifth generation mobile networks is expected to allow deployment of new use cases compared to previous mobile telecommunication standards. This includes massive machine to machine (M2M) \label{M2M} communications in all its variants, including but not limited to device to device (D2D) \label{D2D} , device to cloud, etc. This includes requirements to support stationary devices in the Internet of Things (IoT) \label{IoT}  as well as requirements for highly mobile devices for example in vehicle to everything (V2X\label{V2X}) domain. Power, latency and data rate requirements also vary widely across these different device classes. The introduction of network slices is expected to address these differences in functional requirements. \\

Currently, the migration from 4G to 5G systems is taking place by first deploying 5G solutions in a NSA manner where the first step of the 5G deployment is restricted basically to the new radio aspects of 5G (5G-NR) \label{5G-NR} while the control of the user equipment is still based on 4G protocols, i.e. the core network is still the legacy 4G EPC network. \\

The lack of human interaction results in new trust models that have to be supported. Previously unsolved privacy concerns in 4G are now addressed in the 5G standard. Contrary to the previous generations of standards, the analysis of the security of the 5G system as defined in TS 33.501 \cite{ref1} are already an active concern of researchers before the wide deployment of the standard (cf. \cite{ref2}). Formal analysis of the security procedures by Basin et al. \cite{ref3} has revealed weaknesses that may now potentially still be fixed before 5G standalone (SA) \label{SA} systems are deployed. \\

This report has an objective to cover a full 5G system implementing all features of the standard and not only the radio interface, i.e. a standalone system including the 5G core network (5GC) \label{5GC} architecture. However, given that the reality of the immediate deployments of 5G in the field will actually correspond to NSA deployments (thus phasing in 5G over time), the NSA deployment is also covered in this report. \\

To achieve this, we first present the STRIDE methodology followed in this threat assessment. Then, the following chapter describes the 5G architecture with a focus on the supported security controls. Then the various use cases are analyzed in more detail and the associated security controls to implement them.
Further work is required to validate some of the assumptions made in this document and to check the implementation of various security controls inside both 5G compliant devices and Swiss 5G networks.

\clearpage
\section{Threat Assessment Methodology}
\label{sec1}
The picture below describes the threat assessment methodology in six steps.
\begin{figure}[!htb]
    \centering
    \includegraphics[width=\textwidth]{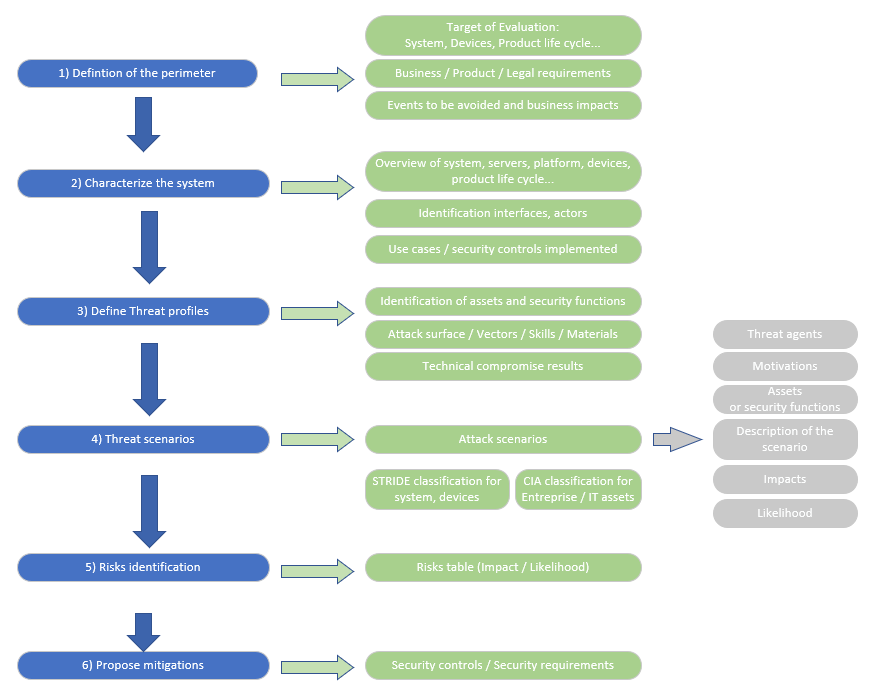}
    \caption{Threat assessement methodology}
    \label{fig1}
\end{figure}

In information security, a threat is a possible danger that might exploit a vulnerability to breach security and therefore cause possible harm. The likelihood of occurrence of a threat is the probability of a successful attack performed by the threat agent in combination with the threat agent’s motivation. \\

The methodology will use a STRIDE classification of threats. The STRIDE methodology for threat analysis requires a formalization of the data flows between the different components.  
STRIDE is a methodology developed originally by Microsoft for describing and categorizing computer security threats. Threats are identified by attacker goal as follows:

\begin{itemize}
    \item \textbf{S}poofing of user or device identity 
    \item \textbf{T}ampering
    \item \textbf{R}epudiation 
    \item \textbf{I}nformation disclosure 
    \item \textbf{D}enial of service  
    \item \textbf{E}levation of privilege 
\end{itemize}

Each component, process, data flow, external entity, and data store – is exposed to a subset of threat categories, as described in the table below. 
Threats on components with STRIDE classification:

\begin{table}[H]
\centering
\resizebox{\textwidth}{!}{%
\begin{tabular}{|c|c|c|c|c|c|c|c|}
\hline
Components                                                                     & Spoofing & Tampering & Repudiation & Information disclosure & \begin{tabular}[c]{@{}c@{}}Denial \\ of \\ Service\end{tabular} & \begin{tabular}[c]{@{}c@{}}Elevation of \\ privileges\end{tabular} & STRIDE \\ \hline
\begin{tabular}[c]{@{}c@{}}External \\ entity or \\ interactors\end{tabular} & X        &           & X           &                         &                                                                   &                                                                      & SR     \\ \hline
Process                                                                        & X        & X         & X           & X                       & X                                                                 & X                                                                    & STRIDE \\ \hline
Data / Keys storage                                                           &          & X         &             & X                       & X                                                                 &                                                                      & TID    \\ \hline
Data flow                                                                      &          & X         &             & X                       & X                                                                 &                                                                      & TID    \\ \hline
Devices                                                                        & X        & X         &             & X                       & X                                                                 & X                                                                    & STIDE \\ \hline
\end{tabular}%
}
\caption{Threats affecting components.}
\label{tab1}
\end{table}

Threats with STRIDE classification and security controls:

\begin{xltabular}{\linewidth}{|M{2cm}|M{9cm}|M{2.5cm}|}
\hline
STRIDE Classification   & Definition                                                                                                                                                                                                                                                                                                                                                                                                                                                                                        & Security Controls       \\ \hline
Spoofing                & - An example of identity spoofing is illegally accessing and then using another user's authentication information, such as username and password.\newline- Spoofing identity of devices could mask fraudulent operations on a system.                                                                                                                                                                                                                                                       & Authentication          \\ \hline
Tampering               & - Data tampering involves the malicious modification of data. Examples include unauthorized changes made to persistent data, such as that held in a database, and the alteration of data as it flows between two computers over an open network, such as the Internet. \newline - Tampering a software, hardware component or keys could impact the security of a system.                                                                                                                 & Integrity Authenticity \\ \hline
Repudiation             & - Repudiation threats are associated with users who deny performing an action without other parties having any way to prove otherwise—for example, a user performs an illegal operation in a system that lacks the ability to trace the prohibited operations.                                                                                                                                                                                                                              & Non-repudiation         \\ \hline
Information disclosure & - Information disclosure threats involve the exposure of information to individuals who are not supposed to have access to it—for example, the ability of users to read a file that they were not granted access to, or the ability of an intruder to read data in transit between two devices, e.g. a mobile phone and an application server. \newline - The ability for someone to extract keys / firmware / hardware IP from devices.                                                  & Confidentiality         \\ \hline
Denial of Service       & - DoS attacks deny service to valid users—for example, by making a server temporarily unavailable or unusable. Protection against certain types of DoS threats could be to improve system availability and reliability.                                                                                                                                                                                                                                                  & Availability Redundancy \\ \hline
Elevation of privileges & - In this type of threat, an unprivileged user gains privileged access and thereby has sufficient access to compromise or destroy the entire system. \newline - Elevation of privilege threats include those situations in which an attacker has effectively penetrated some system defenses and become part of the trusted system itself, a dangerous situation indeed. \newline - Fault attacks on software / hardware components could allow unprivileged access / right on devices. & Authorization           \\ \hline
\caption{Threats definition with stride classification.}
\label{tab2}
\end{xltabular}
\clearpage
\section{5G System Overview}
\label{sec3}
There are several main changes in the 5G architecture compared to the 4G architecture. First, the generic 5G system extends coverage to new frequency spectra that allow to drastically increase data rates and that are well suited for use of massive MIMO (Multiple-Input Multiple-Output) \label{MIMO} and micro-cells. Indeed, transmitters for frequencies in the mm-wave range (including and above Ka bands\label{Ka band}) have intrinsically high directivity thus also providing spatial multiplexing capabilities with more ease than at lower frequencies. Power generation in these frequency ranges is however still difficult and absorption rates by the atmosphere tend to be high. They are therefore unsuitable for macro-cells which are expected to continue to use frequency bands previously allocated for 3G and LTE\label{LTE} networks. \\

Another new frequency range open for use in 5G is around frequencies of 3GHz. Given that this range is in a previously unallocated band (in terms of cellular communications), this frequency range is likely to be used for bandwidth extensions. \\

In stand-alone systems, the 5G architecture also promises much lower latencies to establish a communications channel thus allowing support of use cases as vehicle to vehicle (V2V)\label{V2V} for crash avoidance. Indeed, with higher latencies, some use cases cannot be fulfilled as the introduced delay is prohibitive in safety critical missions. Also, the 5G core network architecture is a service-based architecture. This is reflected in the service based interfaces between network functions implemented on HTTP/2\label{HTTP} over TLS\label{TLS} over TCP/IP\label{TCP}\label{IP}. \\

In non-standalone systems, the core network is still an EPC system. The control plane is handled through an eNB\label{eNB} connection to the mobile equipment while the gNB\label{gNB} only handles user plane traffic. On UE side this implies that two radio interfaces are active. \\

Features that are newly implemented through the core network of the 5G specifications are thus not available in the NSA 5G network. Indeed, the essential difference for this type of 5G system compared to a 4G system is the use of new radio spectra as specified in the 5G-NR part of the specifications. \\

Given that the latency of a connection is partially related to the subframe duration, even a NSA system can already to some extent provide decreased latency. However, the notion of network slices is only available in 5G core systems. Thus, in a NSA deployment, all devices (from low power machine to machine handled in the “massive Machine Type Communications” (mMTC\label{mMTC}) service to an ultra-reliable low latency communications (URLLC\label{URLLC}) device) share access to the radio resources in the same manner. \\

It is noteworthy that the algorithms used for integrity protection and ciphering of NAS\label{NAS} and AS\label{AS} level data did not change compared to the LTE implementation.

\subsection{5G new radio (5G-NR)}
\label{sec3.1}

The 5G radio interface uses the same frequency ranges previously used by LTE solutions and extends them to additional frequency bands. This includes on the one hand frequencies in the sub-6GHz band, particularly newly attributed frequencies around 3.5 GHz and on the other hand frequencies above 6 GHz, more precisely around 24-26 GHz. As in previous mobile communications systems, the various frequency bands are suitable for different use cases. For example, the frequency bands above 6 GHz offer inherently a higher bandwidth (as the bandwidth will to some extent be a value proportional to the center frequency). At the same time, these frequency bands also present higher absorption rates and thus limit the geographical coverage achievable by a single cell. Furthermore, at these frequencies it is getting more complicated to implement near omnidirectional antennas as the antenna to wavelength ratio has the tendency to result in more directive antennas than at lower frequencies. However, given the dimensions of the antenna elements it is also easier to implement beam-forming networks (BFN)\label{BFN} at these frequencies. The frequency bands below 1 GHz still offer the means of achieving coverage with a minimum number of cells (thus achieving coverage in rural areas where the high density deployment of nano-cells would be way too costly).

\subsection{5G non standalone (MR-DC\label{MR-DC} EN-DC)}
\label{sec3.2}

First stages in the 5G deployment focus on the integration of 5G-NR gNBs into the existing 4G system in the context of a multi-radio dual connectivity implementation. This is done by adhering to standard TS 37.340 \cite{ref5} where the system is an instance of an E-UTRA-NR Dual Connectivity (EN-DC)\label{EN-DC} deployment (cf. 4.1.2 of \cite{ref5}). This means that the core network is still the 4G EPC and the master nodes of the dual connectivity are 4G eNBs. The 5G base station is integrated as an en-gNB\label{en-gNB} into the system and acts as a secondary node. It only exchanges user plane data with the core network. All control data is exchanged with the eNB over the X2 link. From a UE \label{UE} perspective, the control plane is located in the eNB while user plane data are transmitted over the gNB. This dual connectivity system also implies that the UEs that support this mode have to integrate concurrent 4G and 5G radio interface support. In terms of power consumption this is naturally not ideal as two radio frontends have to be powered up in parallel. Therefore, this mode might be unsuitable for low-power applications in the IoT context. \\

Finally, UEs supporting this mode of operation have to use the standard 4G network attach procedures which implies sending their IMSI\label{IMSI} in clear to the network during the first attach. This means that the identity concealment feature introduced for 5G is not usable in the non standalone deployments and IMSI catching is still possible without any increased difficulty. 

\begin{figure}[H]
    \centering
    \includegraphics[width=\textwidth]{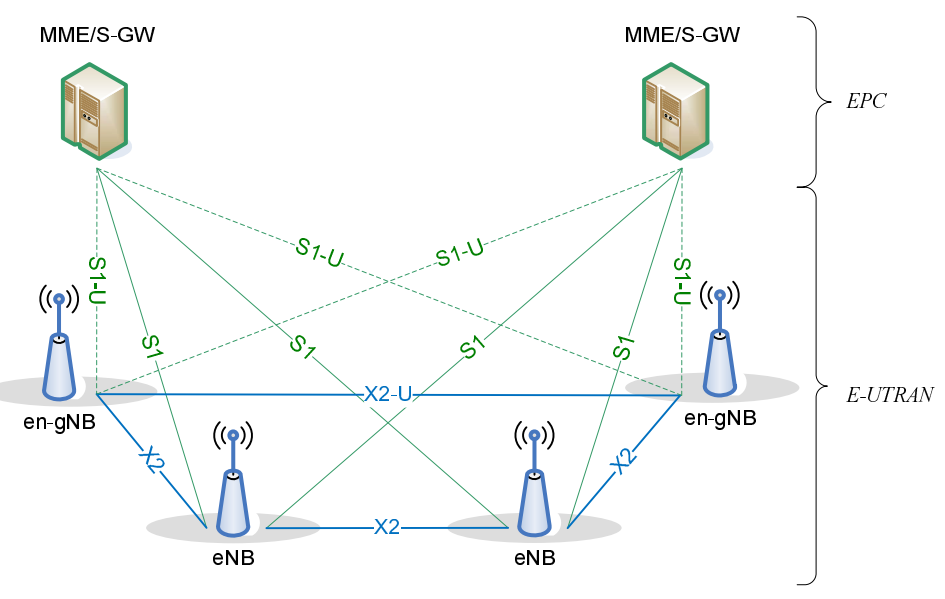}
    \caption{NSA 5G network according to \cite{ref5}}
    \label{fig2}
\end{figure}

\subsection{5G standalone}
\label{sec3.3}

In the case of a standalone 5G deployment (or of a dual connectivity deployment using a 5G core network), not only the radio interface is different from the previous 4G system. The whole core network is also different. In 5G, the architecture has been designed in a way to achieve a cleaner separation of the control and user planes. The core network has also been redesigned using a service based architecture. In a way, this makes virtualization of some network functions easier. Once virtualized, the network functions can naturally also be implemented as cloud instances. To guarantee security of virtualized network functions, the operator of the 5G system has to pay attention to the isolation mechanisms between the virtual machines implementing them. Also, the implicit level of trust in a serving network has been reduced and some new security features have been implemented. Authentication and access management functions are now in two different building blocks of the system. \\

\begin{figure}[H]
    \centering
    \includegraphics[width=\textwidth]{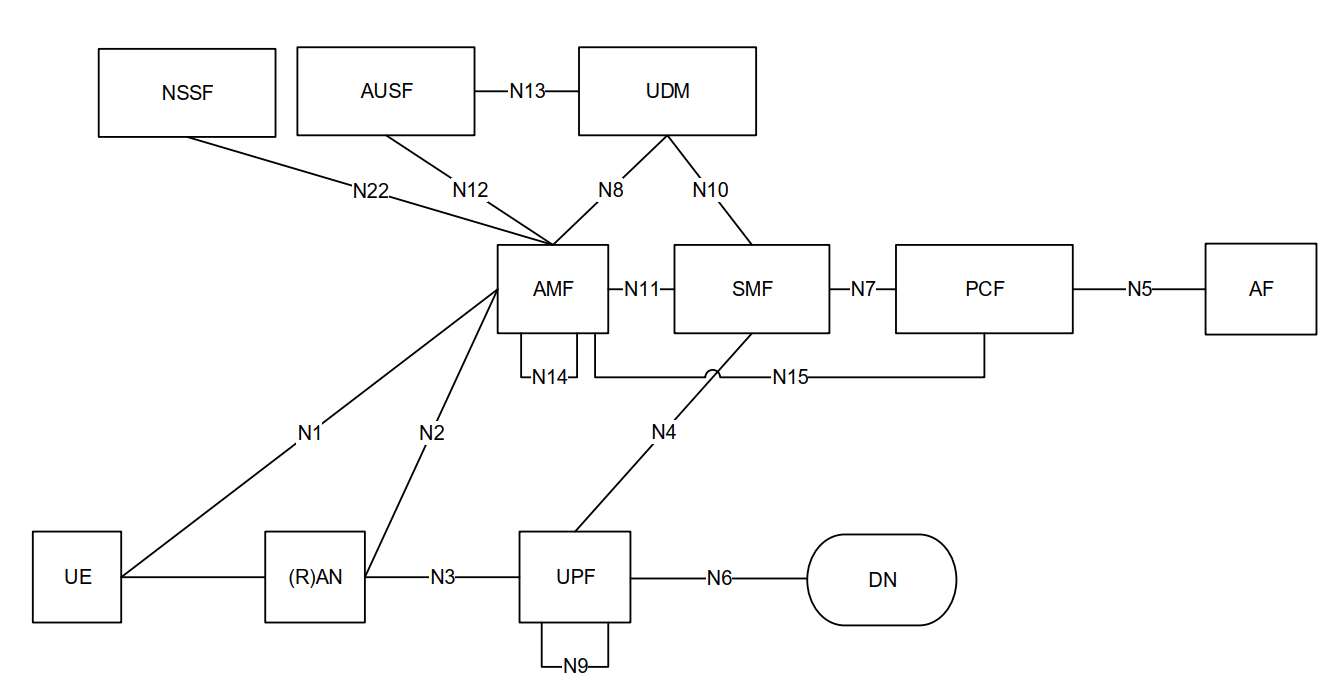}
    \caption{Reference architecture of the 5G system in a non-roaming context \cite{ref7}}
    \label{fig_5G_NS}
\end{figure}

Figure \ref{fig_5G_NS} shows the various reference points of the 5G system architecture if no roaming is involved, i.e. if the serving network corresponds to the home network. As can be seen, the access and mobility management function (AMF)\label{AMF} are clearly separated from the session management function (SMF)\label{SMF}. The unified data management (UDM)\label{UDM} of the home network and the USIM\label{USIM} of the UE contain the same long-term keys used for further key derivation during the authentication process. The authentication server function (AUSF)\label{AUSF} is located in the home network of the device and performs its authentication. It also provides high level keys to the AMF that initiated the authentication session.  

\begin{figure}[H]
    \centering
    \includegraphics[width=\textwidth]{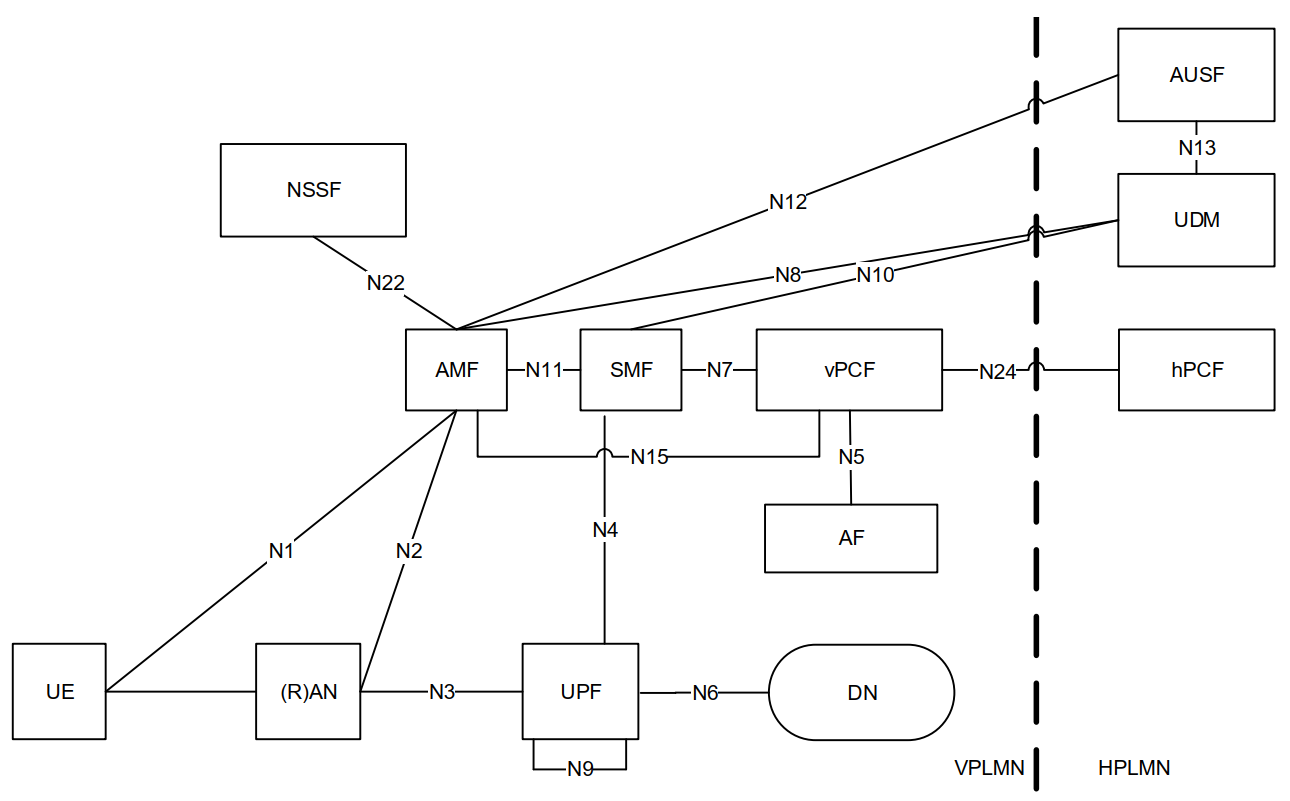}
    \caption{Roaming situation with local data breakout \cite{ref7}}
    \label{Roaming_1}
\end{figure}

\begin{figure}[H]
    \centering
    \includegraphics[width=\textwidth]{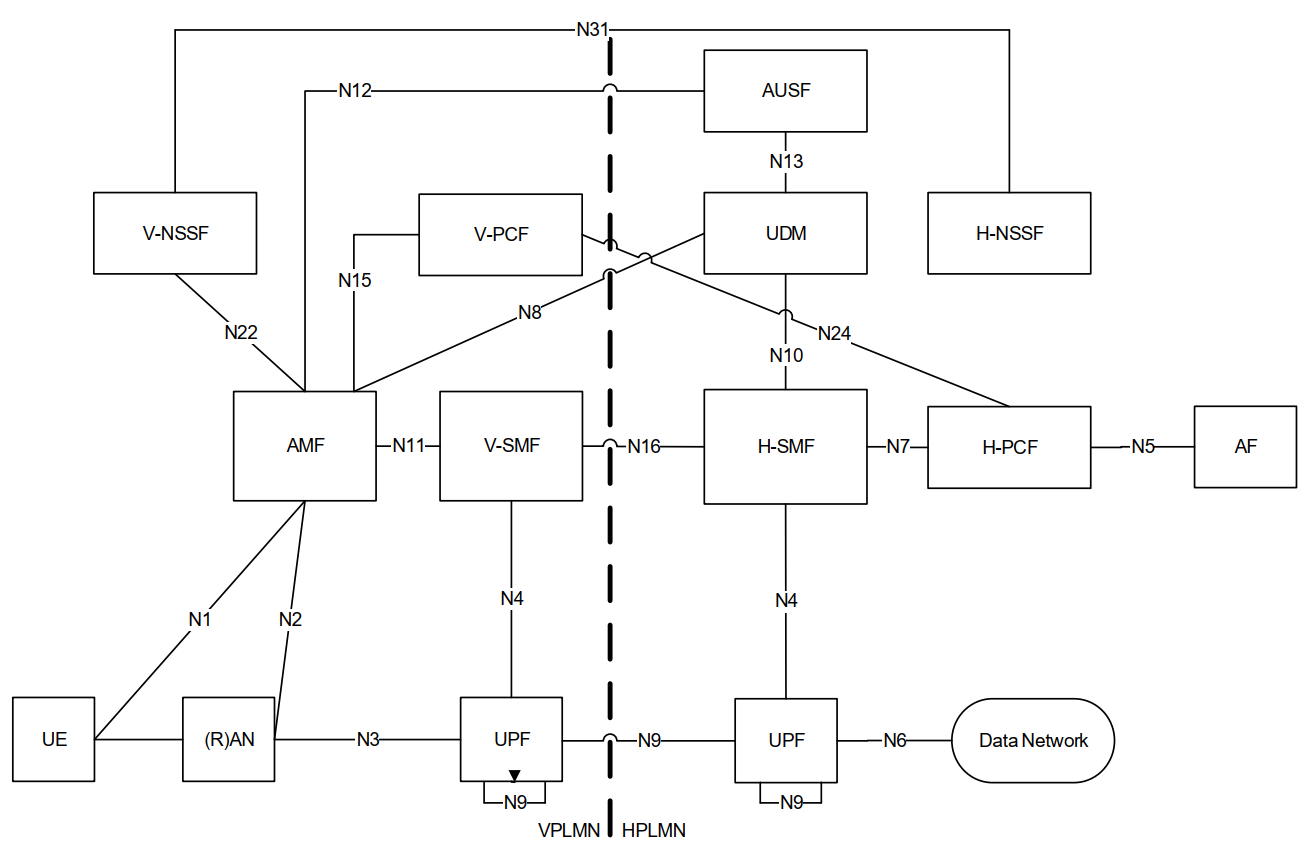}
    \caption{Roaming situation with data routing through the home network \cite{ref7}}
    \label{Roaming_2}
\end{figure}

In case of roaming, there are two possibilities. First, the data can be routed through the data network of the visiting network (cf. Figure \ref{Roaming_1}) or, second, data can be routed through the data network of the home network (cf. Figure \ref{Roaming_2}). In both figures, all control plane traffic exchanged between the visiting and home network is actually passed through the security edge protection proxies (SEPP)\label{SEPP} of the visiting and home networks. The AMF contains a security anchor functionality (SEAF)\label{SEAF} that holds the keys provided to it by the AUSF during the authentication procedure of the UE to the network. \\

When the UE first registers to the system, the core network is not yet in the possession of a valid security context and thus, the first exchanges between the UE and core network are in clear and not integrity protected. However, in the case of a full 5G system, the USIM of the device also contains the public key of its home network. This enables the UE to conceal its identity to anyone not holding the corresponding private key, i.e. to anyone other than the UDM of the home network. However, it already has to communicate the identity of its home network to the serving network. This home network identifier is communicated in clear. 

\subsection{New security features in 5G compared to 4G}
\label{sec3.4}

In 5G standalone implementations, some new security features have been implemented in order to mitigate previously identified vulnerabilities. At the same time, previously existing security controls are in general still present in 5G. \\
Contrary to previous versions of 3GPP \label{3GPP} standards, the UICC\label{UICC} of the UE now contains an asymmetric key element, the public key of the home network for use in elliptic curve algorithms. The (limited) use of asymmetric cryptographic algorithms allows the transmission of confidentiality protected information to the core network without a previous key negotiation with this network. This mechanism avoids previous IMSI catcher attacks that allow tracking a mobile phone as the unprotected IMSI in the initial attach request has been replaced by an obfuscated SUCI\label{SUCI} in the initial registration request. As the other permanent identifiers allowing to directly map a user to a connection.

\begin{table}[H]
\begin{tabular}{|M{4cm}|M{5cm}|M{5cm}|}
\hline
Security feature                                                                                       & Applies to 4G                               & Applies to 5G                                 \\ \hline
IMSI obfuscation on radio link                                                                         & No                                          & Yes, using ECIES\label{ECIES} scheme                       \\ \hline
User plane encryption on radio interface level                                                        & Yes (operator choice)                       & Yes (operator choice)                         \\ \hline
User plane integrity protection on radio interface level                                              & No                                          & Yes (operator choice)                         \\ \hline
RRC\label{RRC} message integrity protection                                                                       & Yes, EIA0 only allowed for emergency calls & Yes, NIA0 only allowed for emergency calls   \\ \hline
RRC message encryption                                                                                 & Yes (operator choice)                       & Yes (operator choice)                         \\ \hline
NAS message integrity protection                                                                       & Yes, EIA0 only allowed for emergency calls & Yes, NIA0 only allowed for emergency calls   \\ \hline
NAS message encryption                                                                                 & Yes (operator choice)                       & Yes (operator choice)                         \\ \hline
Authentication of UE to serving network                                                                & Yes                                         & Yes                                           \\ \hline
Authentication of UE to home network even if using untrusted serving network                          & No                                          & Yes                                           \\ \hline
Network slicing to provide differentiated handling of service requirements for different applications & No                                          & Yes                                           \\ \hline
\end{tabular}
\caption{Comparison of security features in 4G and 5G.}
\label{tab3}
\end{table}

Naturally, the messages can only be protected either in integrity or confidentiality if a security context has been established between the UE and the network. This means that the previous attacks on the attach request/attach reject procedure (e.g. reject cause 8, i.e. EPS\label{EPS} and non-EPS services not allowed, as described in subsection 5.5.1.2.5 of TS 24.301 \cite{ref13}) are still applicable in the comparable registration request/registration reject procedure (e.g. reject cause 3, i.e. illegal UE, as described in subsection 5.5.1.2.5 of TS 24.501 \cite{ref11}). However, it can be noted that reject cause 8 is no longer supported in the 5G registration procedure. \\


So the 5G standalone is not perfect but it is more secure against IMSI catchers than previous protocols if the operator implements all 3GPP recommendations.
\clearpage
\section{5G System Details}
\label{sec4}

\subsection{Involved entities}
\label{sec4.1}

\begin{figure}[H]
    \centering
    \includegraphics[width=\textwidth]{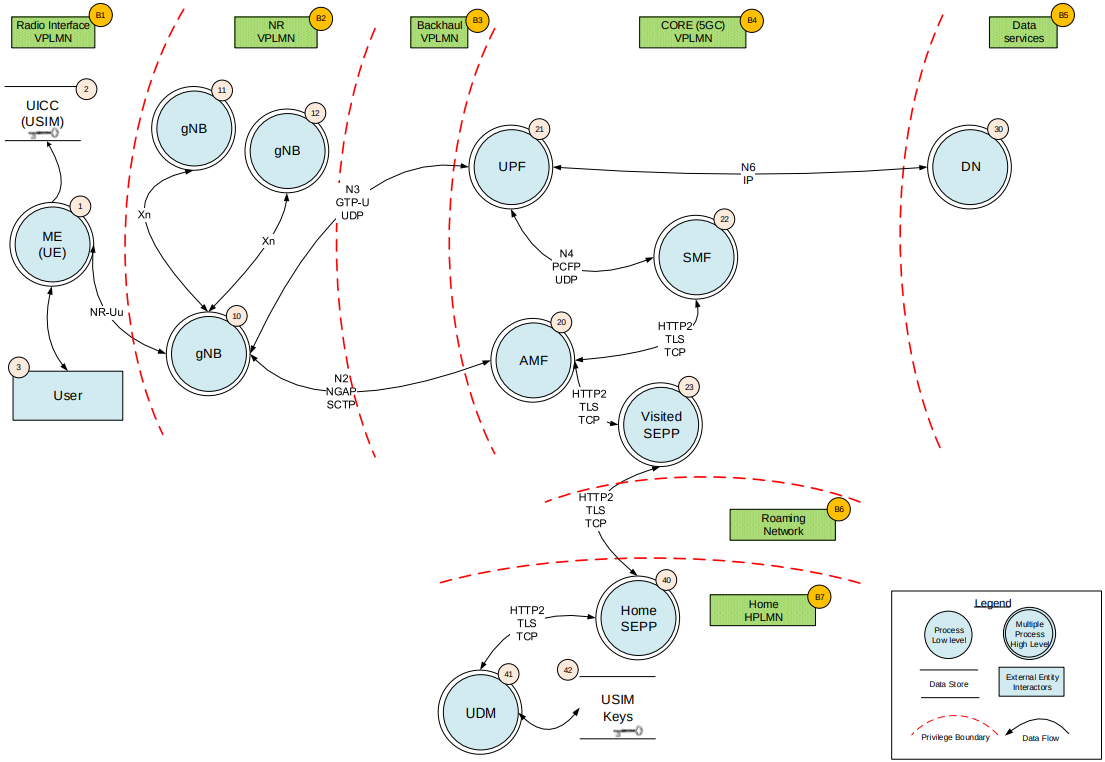}
    \caption{Typical data flow diagram in the absence of IPsec\label{IPsec}} 
    \label{fig6}
    \label{NGAP}
    \label{UDP}
\end{figure}

The following entities are involved in the communications between a UE and a terminating endpoint in the packet data network. Figure \ref{fig6} on page~\pageref{fig6} shows the data flow between these entities in case of a roaming situation. If the UE connects directly to the home network, the AMF and SMF connect directly to the UDM. In some machine type communications, the user is not actually part of the data flow.

\subsubsection{UE}
\label{sec4.1.1}

In this report, user equipment is understood to not only include classical user terminals that offer standard data or voice services to the user but also devices that are used for M2M communications. This naturally impacts on how various device states can be handled and thus also on the impact of an attack. For example, if an attacker can place the UE in a ROAMING NOT ALLOWED state for a given PLMN\label{PLMN}, then the standard often provisions that the UE shall not try to register again to this PLMN except upon a power cycle. A user might get aware that a hand-held device is not connected to any network and try to do a power off/on cycle. However, for a device used for M2M communications there might not be any user interaction after the initial deployment of the device. Thus, if this network lock-out situation is not handled in an application level state machine of the M2M device, then the device might remain permanently deregistered from the network even once the attacker has stopped the active interference with the serving network. \\

The UE is the combination of the mobile equipment (ME)\label{ME} and the UICC that holds the subscription details of the user including the root keys used for all further integrity and confidentiality protection mechanisms. In 5G, the ME can decide to obfuscate the unique subscription identifier, the SUPI\label{SUPI}, by applying an ECIES protection scheme and only transmitting the SUCI. The preferred network slice and the home network name still are transmitted in clear even when using this obfuscated subscription identifier. Knowledge of the preferred network slice might reveal information about the device class and knowledge of the home network still reveals a limited amount of personal information. 

\begin{table}[H]
\begin{tabular}{|M{3cm}|M{3cm}|M{8cm}|}
\hline
UE identifier                     & Permanent          & Context                                                                                                                                                                                       \\ \hline
SUPI (equals IMSI in 3GPP access) & Yes                & Stored in UICC inside the UE and in the UDM, used to identify the UE between network functions ; only transmitted in emergency call registration                                             \\ \hline
SUCI                              & No, only used once & Transmitted on radio level before security context establishment and before allocation of 5G-GUTI; applied protection profile decided by the UE; only the home network can extract the SUPI \\ \hline
PEI\label{PEI}                    & Yes                & ME identifier, transmitted after security mode setup and thus potentially confidentiality protected                                                                                           \\ \hline
5G-GUTI                           & Semi permanent     & Allocated at the end of the registration procedure and sent to the UE after security mode setup. Later used in clear to identify a security context to be used between UE and 5GC / RAN\label{RAN}\\ \hline
S-TMSI                            & Semi permanent     & Short version of 5G-GUTI used in paging procedure                                                                                                                                             \\ \hline
C-RNTI                            & Short term         & Used as a UE identifier for RRC connections and scheduling                                                                                                                                    \\ \hline
\end{tabular}
\caption{Various identifiers for the UE.}
\label{tab4}
\end{table}

\paragraph{Battery powered sensor devices}

The various types of UE also have different characteristics in terms of power supply, data rates and required QoS\label{QoS} levels in terms of latency. Indeed, a sensor device might be battery powered and in a fixed location. The device itself does not require to transmit huge amounts of data but there could be a huge number of this kind of device being covered by a single gNB. Given its stationary nature, cell switches are only to be expected if the radio network is reconfigured. In a default 5GC network, this device family is expected to be handled in the mMTC slice of the network. In this type of slice, the core network might want to only allocate short data slots to the devices but do so on a regular basis. In this way, the latency of device communications might be relatively high, but a single cell can handle a much bigger number of devices compared to a standard enhanced mobile broadband (eMBB\label{eMBB}) slice.

\paragraph{Externally powered mobile devices}

Another device category could be the advanced driver assistance system inside a vehicle. The device being directly connected to the power supply of the vehicle, low power considerations to increase the lifetime of the device are no longer to be considered. However, this device is now highly mobile and it will change cells frequently. Also, in the context of autonomous driving, the device might have to exchange data reliably and constantly with other vehicles. For parts of the driver assistant system, such as collision avoidance sub-systems, the amount of data to be exchanged with other vehicles or other edge devices might once again be limited but the data has to be transmitted as soon as possible, i.e. the latency requirements of this sub-system might be superior to normal communications contexts. These requirements can be handled in a correct URLLC slice implementation. \\

Map sub-systems might have more standard requirements of punctual high data rate connections to a map update server. This sub-system inside an autonomous vehicle might therefore request to be registered in an eMBB slice. \\

In all use cases, the UE contains a USIM (either a distinct hardware element or directly embedded in the main chipset) that stores and handles the long-term key shared with the home network. This long-term key is at the basis of all key derivations. On operator side, the same key is stored in the UDM. 

\subsubsection{gNB}
\label{sec4.1.2}

The gNB is responsible for the implementation of the radio interface with the UE. It is then splitting the data into control and user plane segments and sends them accordingly to the various end points. \\

In a roaming use case, the gNB is part of the serving network that is different from the home network to which the device has subscribed. It holds “low-level” keys shared with the UE to protect both in integrity and confidentiality the RRC and user plane exchanges with the UE. While highly recommended by the standard, the network has the choice to enable or not RRC and user data encryption between the gNB and the UE. \\

\begin{figure}[H]
    \centering
    \includegraphics[width=5cm]{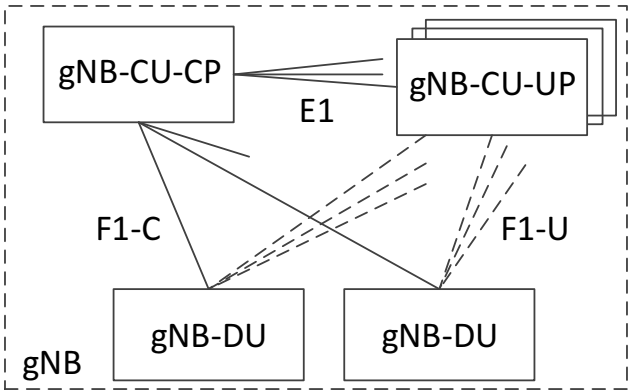}
    \caption{Architecture of a split of the gNB into DU\label{DU} and CU\label{CU} and split of the central unit in control plane and user plane entities according to Figure 6.1.2-1 of \cite{ref14}}
    \label{fig_dist_unit}
\end{figure}

It can be split into a central unit - distributed unit architecture (cf. Figure \ref{fig_dist_unit}) where the distributed units include the physical radio interface. In the below example, the central unit is further split into a function handling the control plane and several functions handling the user planes. As specified in section 9.8 of TS 33.501 \cite{ref1}, the control interfaces E1 and F1-C shall be integrity, replay and confidentiality protected using IPsec and DTLS while the traffic on the F1-U interface shall be protected using IPsec. However, as is quite common, the specification also states that the use of cryptographic solutions is an operator's choice and is not required if the gNB is placed in a physically secured environment.

The gNB requires some manual configuration to know the addresses of the AMFs it can communicate with.

\subsubsection{AMF}
\label{sec4.1.3}

The AMF is the endpoint for NAS messages from the UE on the N1 interface. It is located in the serving network. On control plane level, it communicates with the access network, i.e. the gNBs on N2 interface. The AMF also contains the SEAF that is responsible for the interactions with the home network during the authentication of a UE.

\subsubsection{AUSF}
\label{sec4.1.4}

The AUSF is responsible for the authentication of UEs with the network. It is located in the home network and communicates on the one hand with the UDM to retrieve long-term keys and on the other hand with the AMF to which it communicates keys derived from the keys recovered from the UDM and the other input parameters to its key derivation function. It also sends the SUPI associated to an authentication session to the corresponding SEAF inside the AMF at the end of a successful authentication. 

\subsubsection{SEPP}
\label{sec4.1.5}

The security edge protection proxy is responsible for filtering and transmitting messages between serving and home networks. It hides its internal network topology from the other networks and is capable of authenticating the other SEPP it communicates with.

\subsubsection{UDM}
\label{sec4.1.6}

The UDM contains the long-term keys and associated shared secrets used inside the authentication and key derivation process for subscribers to this network. It also contains the private key used to obtain the SUPI from the concealed SUCI. \\
\noindent
The UDM also contains the subscription details of a given UE and holds the identity of the network functions currently serving the UE.

\subsubsection{SMF}
\label{sec4.1.7}

The session management function (SMF) performs the selection of the appropriate user plane functions (UPF)\label{UPF} for a given data session of a UE with the network. It also provides QoS parameters to the UPF and provides a monitoring and control interface between the user and control planes. \\
\noindent
It allocates the IP addresses to the UEs and is responsible for downlink data notification.

\subsubsection{UPF}
\label{sec4.1.8}

The user plane function is the PDU\label{PDU} session point of interconnect to the data network. It handles the QoS for the user plane, buffers downlink data and routes and forwards packets. In the context of lawful intercept, UP collection takes place in the UPF.

\subsubsection{Other network functions}
\label{sec4.1.9}

Other less directly exposed network functions are the Policy Control Function (PCF)\label{PCF}, the Unified Data Repository (UDR)\label{UDR}, the Network Exposure Function (NEF)\label{NEF}, the Network Slice Selector Function (NSSF)\label{NSSF} and the Network Repository Function (NRF)\label{NRF}. They handle purely service based interfaces.

\paragraph{PCF}

The PCF provides policy controls for service data flows and PDU sessions. Its services (cf. subsection 5.2.5 of TS 23.502 \cite{ref15}) are used by the AMF, the SMF, the NEF and the PCF of the visited network in case of roaming. 

\paragraph{UDR}

The UDR can be composed of one or several instances. Besides other storage, it provides storage for subscription data used by the UDM and for policy data used by the PCF.

\paragraph{NEF}

Application functions outside the core network can still access some information. However, the NEF decides which level of information (capabilities and events) shall be exposed to the external AF. It can be considered a sort of firewall that will only serve requests from the application functions outside the 5GC's trust boundary that pass its rules. 

\paragraph{NSSF}

During the registration procedure and potentially also during PDU service requests, the UE will request a list of network slices. The NSSF provides the mapping of requested to supported network slices and might also trigger a change in AMF associated to a UE. 

\paragraph{NRF}

The NRF maintains a repository of network functions inside its core network. This repository also contains configuration data of these NFs such as their location, associated network slice and many more. A network function can discover other network functions that it has to connect to in order to offer its services through the discovery service of the NRF. It also serves as authorization server based on the OAuth2 authorization scheme and provides access tokens to the authorized client. The NRF is only used if the topology of the core network is not configured statically. In this report, it is not shown in most of the data flow diagrams as it is not part of the data flow during a data session between the user and the data network. However, it allows to the various network functions in the core network to discover the network topology and provide service authorization and thus interconnect with each other. 

\subsection{Cryptographic algorithms and protection schemes}
\label{sec4.2}

The 3GPP standards specify four pairs of algorithms that are used to ensure the confidentiality and integrity of communications between the UE and the RAN and the AMF respectively. It also specifies three encryption schemes to obfuscate the SUPI. Finally, it specifies the use of IPsec and (D)TLS for some of the communications between the gNBs and the 5GC or between entities of the 5GC. The encryption and integrity protection algorithms used by the UE are the same in 5G as in 4G. NEA\label{NEA} (5G Encryption Algorithm) is a ciphering algorithm; NIA\label{NIA} (5G Integrity Algorithm) is an integrity algorithm. How the various interfaces are protected is summarized in Table 5 below.

\begin{table}[H]
\begin{tabular}{|M{4.5cm}|M{4cm}|M{5cm}|}
\hline
Interface                       & Protected against                       & Protection mechanism                                                                                        \\ \hline
N2, N3, Xn, E1, F1              & Confidentiality, integrity, anti-replay & Operator choice to apply IPsec ESP\label{ESP} and IKEv2\label{IKEv2} certificates-based authentication      \\ \hline
SUPI/SUCI on NR-Uu, N1          & Confidentiality                         & ECIES                                                                                                       \\ \hline
PEI on N1                       & Confidentiality, integrity              & Integrity mandatory through NIA1-NIA3, confidentiality operator choice through NEA1-NEA3                    \\ \hline
Signaling data on NR-Uu and N1  & Confidentiality, integrity              & Integrity mandatory through NIA1-NIA3, confidentiality operator choice through NEA1-NEA3                    \\ \hline
User plane data on NR-Uu        & Confidentiality, integrity              & Operator choice to apply integrity/confidentiality protection through NIA1-NIA3 and NEA1-NEA3 respectively  \\ \hline
Service based interfaces        & Confidentiality, integrity, anti-replay & Operator choice to apply TLS                                                                                \\ \hline
\end{tabular}
\caption{Protection mechanisms for the 5g interfaces.}
\label{tab5}
\end{table}

\subsubsection{NEA0 / NIA0}
\label{sec4.2.1}

The first pair of cryptographic algorithms are «null algorithms» that do not provide encryption nor integrity protection. These algorithms are supposed to be used in scenarios where it is impossible or not required to authenticate the UE. They can be used for emergency calls, for instance. However, apart from this context, the standard clearly excludes them from the list of allowed algorithms. Note that the null integrity protection algorithm does not provide any proof of message integrity but it does increase an unnecessary overhead to the message. For short messages (e.g. in the context of massive machine type communications), the added overhead can be significant compared to the useful message length. \\

The SUPI can also be \say{transformed} into the SUCI using the null-scheme. According to the specifications this scheme shall only be used if the operator choses to not provision the USIM with its public key or enforces the use of this scheme (p. 82 in \cite{ref1}). However, some devices seem to also chose the null-scheme themselves (e.g. some Wi-Fi to 5G access points).

\subsubsection{128-NEA1 / 128-NIA1}
\label{sec4.2.2}

The second pair 128-NEA1/128-NIA1 is based on the Snow 3G stream cipher. Snow 3G is a 32-bit word-oriented stream cipher supporting 128-bit keys, which was also part of the 3G standard. The 3GPP standard supports the encryption (128-NEA1) or authentication (128-NIA1) of blocks of data from 1 to 20’000 bits.

\subsubsection{128-NEA2 / 128-NIA2}
\label{sec4.2.3}

The third pair 128-NEA2/128-NIA2 is based on the AES\label{AES} block cipher. AES is probably the block cipher that has been the most analyzed in the history of cryptography, and its security seems unquestionable, even in a 5G context.

\subsubsection{128-NEA3 / 128-NIA3}
\label{sec4.2.4}

The fourth pair 128-NEA3/128-NIA3 is based on the ZUC\label{ZUC} stream cipher. ZUC is the most recent cipher out of the three that has been designed by the Data Assurance and Communication Security Research Center of the Chinese Academy of Sciences. It is the least analyzed standard 5G cipher, although no significant attack is currently known.

\subsubsection{Elliptic curve integrated encryption scheme (ECIES)}
\label{sec4.2.5}

The SUPI obfuscation mechanism can be achieved using two elliptic curve based protection schemes. Both schemes use AES-128 in counter mode for the symmetric encryption of the of the data and either the curve Curve25519 or secp256r1 for the ECDH\label{ECDH} key exchange. The use of this encryption scheme is new in 5G. However, it is currently only used for the SUPI to SUCI transformation and the home network's public key is not used in the authentication of any other message or parameter.

\subsubsection{IPsec}
\label{sec4.2.6}

For non service based interfaces between the (R)AN and the 5GC and inside the 5GC, the connection is expected to be secured using IPsec with protection profiles defined in chapters 4 and 5 of TS 33.210 \cite{ref10}.

\subsubsection{(D)TLS}
\label{sec4.2.7}

Among other interfaces, all service based interfaces shall be protected using TLS. The specifications require a minimum level of robustness of the cryptographic suites used in the TLS implementations inside the 5G system. Within the TLS profiles allowed for 5G, the authorized ciphersuites present all AEAD (authenticated encryption with associated data)\label{AEAD} features. Between network functions, the use TLS without encryption is not allowed. The recommended cryptographic algorithms for symmetric encryption are AES-128 or even AES 256 while ECDH implementations have a minimum key length of 255 bits and DHE\label{DHE} implementations have a minimum key length of 2048 bits (and key length of 4096 bits have to be supported).

\subsection{Processes}
\label{sec4.3}

The main processes that are expected to be the surface of choice for attacks are the registration of a device to the core network (and its subsequent re-registration) and the access stratum initialization / RRC configuration. The authentication procedure of the UE with its home network is also part of the initial registration of the device to a network.

\subsubsection{Registration to the core network}
\label{sec4.3.1}

The registration to the core network is triggered by a UE initiated procedure. The sequence of messages is shown in Figure 8. During the initial registration, the UE does not yet hold a 5G-GUTI \label{5G-GUTI}  or a valid security context. The same applies to the 5GC. In previous 3GPP standards, the UE used its IMSI to identify itself to the core network. This raised however serious privacy concerns and lead to the 5G standard introducing the concealed SUPI (corresponding to the IMSI in previous versions), the SUCI. The SUCI contains the SUPI and a random nonce that are encrypted using the home network’s public key. It contains also the identifier of the home network in clear in order for the serving network to be able to forward the data for decryption to the right entity. For an external attacker who does not have access to the database inside the UDM establishing a relationship between the SUPI and the subscriber, the knowledge of the home network identifier sent in clear during the initial registration request and of the PEI / IMEI\label{IMEI} of the device provide the same level of information as the IMSI (as provided by 4G IMSI catchers).

\begin{figure}[H]
    \centering
    \includegraphics[width=\textwidth]{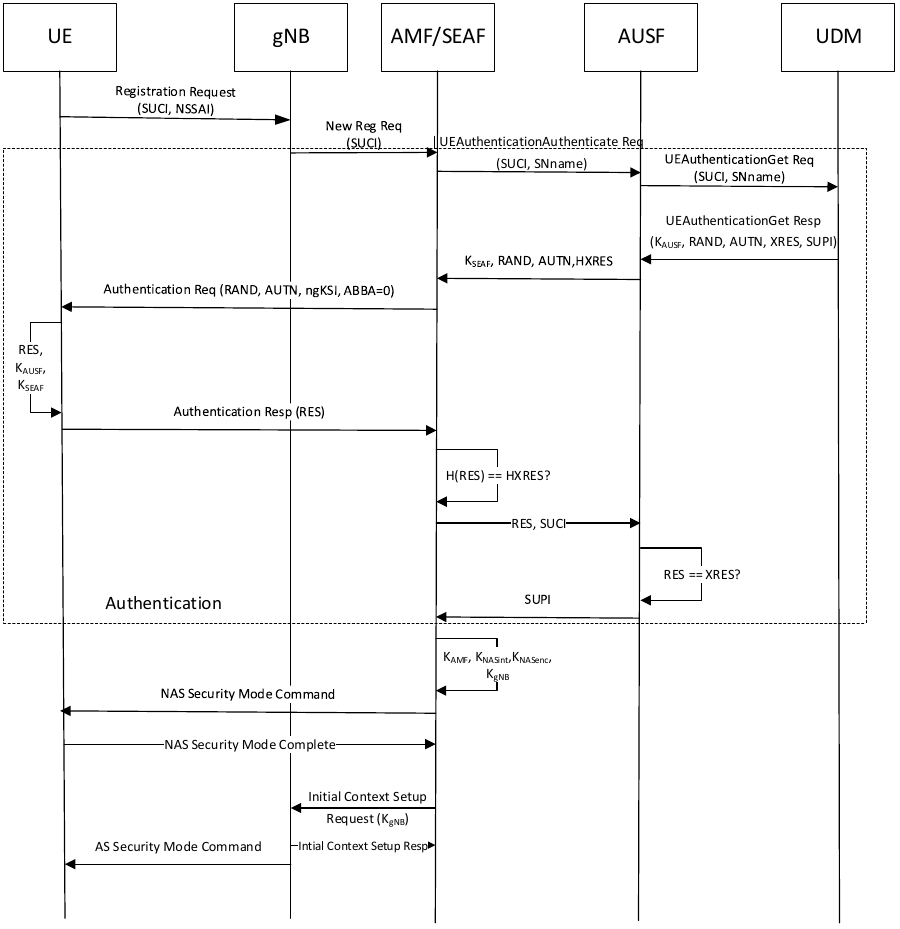}
    \caption{Initial device registration and authentication (until full security mode setup)}
    \label{fig_init_reg}
\end{figure}

\paragraph{Initial registration request}

In an initial registration request to a network (cf. Figure \ref{fig_init_reg}), the UE sends the registration request containing the SUCI and the preferred network slice indication to the gNB. The gNB selects the right AMF (based on the network slice indicator) to serve future NAS messages. The AMF then initiates the authentication procedure for this UE. 

\paragraph{Authentication} 

The AMF initiates the authentication procedure by sending the authentication request for this UE to the AUSF of the home network of the UE. This authentication request contains the SUCI of the UE to be authenticated and the name of the serving network. The AUSF then transmits the same information to its UDM. The UDM can decrypt the SUCI and thus obtains the SUPI of the UE to be authenticated. The UDM being in possession of the same pre-shared secrets as the USIM of the UE, it first draws a random number to serve as input to its authentication challenge. It then computes the expected result of the challenge response XRES and the authentication token for this challenge AUTN. It also computes the $K_{AUSF}$ that will be used by the AUSF for further key derivation for the security context associated to this authentication session. All these data elements are returned to the AUSF. \\

In 5G networks, the home network does not have the same level of trust in the serving network as in itself. This is an innovation compared to 4G implementations where the MME\label{MME} in the visited network directly checked the result of the authentication response from the UE. To reduce the implicit level of trust into the serving network, the AUSF computes the hash of the XRES and replaces the XRES in the authentication vector with the computed HXRES. It also derives the key $K_{SEAF}$ that will be used as input by the serving network for the derivation of the AMF key once the authentication has been successful. The AUSF then transmits the modified authentication vector and $K_{SEAF}$ to the SEAF inside the AMF that initiated the authentication session. \\

At this moment in time, the AMF does not yet have all elements to construct the network side security context. It will send an Authentication Request message on NAS level to the UE (using the N1 interface). In the current version of the protocol, this message contains an Anti-Bidding down Between Architectures (ABBA)\label{ABBA} parameter of all zeros. It further contains the challenge RAND and the authentication token AUTN from the authentication vector. Finally, it contains the expected 5G key set identifier ngKSI to be used for this session by the UE if the authentication is successful. \\

On reception of the Authentication Request message from the AMF, the UE will request its USIM to check the freshness of the AUTN token. On success, it computes the RES challenge response based on the received RAND challenge and transmits its response to the AMF. The AMF can check the success of the authentication by applying the same hash function on the RES value as previously applied by the AUSF on the XRES value. \\

On success of the HRES comparison by the AMF, the AMF can then send the RES value to the AUSF that independently verifies that the challenge response corresponds to the expected result. Given that the AMF did not know the challenge response prior to its reception from the UE, the AUSF can know that the AMF does not tamper the response of the UE. At the end of the successful authentication, the AUSF transmits the SUPI to the AMF. 

\paragraph{Final registration steps}

The SUPI is required for the derivation of the key $K_{AMF}$ used as input for $K_{NASint}$, $K_{NASenc}$ and $K_{gNB}$. With knowledge of these keys, the AMF can activate security on NAS level with the UE and provide the gNB with means to do the same. \\

The AMF first configures the security mode on NAS level with the UE by sending the NAS Security Mode Command to the UE. On acknowledgement of the successful NAS level security mode configuration by the UE through an integrity protected NAS Security Mode Complete message, the AMF sends the InitialContext Setup message to the gNB. This message contains the key $K_{gNB}$. \\

This key can be used by the gNB to derive the complete set of keys used for the integrity and confidentiality protection both of signaling and user data on access stratum level, i.e. $K_{RRCint}$, $K_{RRCenc}$, $K_{UPint}$ and $K_{UPenc}$. \\

As shown at the end of the sequence in Figure \ref{fig_init_reg}, the gNB first acknowledges the InitialContext Setup to the AMF before activating the security mode with the UE. \\

From this moment in time, all further exchanges between the network and the UE are at least integrity protected on both AS and NAS levels.

\subsubsection{Access stratum initialization/radio resource configuration}
\label{sec4.3.2}

On AS side, the gNB uses keys derived from the key $K_{gNB}$ to integrity protect and potentially encrypt data both on control plane and on user plane levels. Except for emergency call procedures, all RRC messages are expected to be integrity protected (and not using the NULL integrity protection scheme NIA0) once the AS security context has been established. \\

Integrity protection on user plane level is optional and controlled from the network side. Given that integrity protection adds some overhead, this increase in overall data rate may be deemed unacceptable, e.g. in some mMTC use cases with low data rates. \\

Encryption of control and/or user plane data is also only optional on gNB level even though the standard highly recommends its activation. However, as a network can choose to not activate ciphering of data, the UE will have to support its absence. The activation of user data confidentiality and integrity protection is nominally based on the security policy sent by the SMF to the gNB.

\subsubsection{Authentication and authorization between network functions}
\label{sec4.3.3}

In the service base architecture, the different network functions can be virtualized, implemented as cloud instances in different virtual machines and in general not be in the same protected physical domain. \\

In order to avoid information disclosure and misconfiguration of the services, a service consumer has to present proof of being authorized to use the service and the consumer will at the same time require to know that it communicates with the authentic service producer. In the 5GC, a new network function instance can first discover the network functions it is authorized to access by authenticating itself to the appropriate NRF. The NRF can also serve as authorization server. In this case it will provide the consumer network function with an authorization token for the allowed services of the service producer. \\

To make certain that the token is not intercepted by a man in the middle and sent to the authentic service producer, the service consumer network function first authenticates the service producer network function by establishing a TLS session on transport layer level. Once established, the TLS session protects the exchanged data in confidentiality and authenticity. The service consumer network function then proves its access rights through presentation of the authorization token to the producer network function. \\

As an alternative to this token-based authorization, section 13.3.2 of TS 33.501 \cite{ref1} allows to use mutual authentication of the two network functions. In this case, the service producer network function is still required to verify the authorization of the service consumer NF\label{NF}.

\subsection{Summary of differences between 4G and 5G}
\label{sec4.4}

Naturally the most visible difference between 4G and 5G is the extension of the frequency bands to mm-waves. However, from a security point of view, this change has no direct impact except potentially increasing the difficulty to jam the entire frequency range covered by a 5G UE. At the same time, the higher frequencies are providing coverage in nano-cells. The increased importance of beamforming might however be considered an advantage of the 5G system. Indeed, it is at least in theory possible that the beamforming property of a gNB is used to cancel the direction of arrival of a physical jammer. This would at least reduce the impact of the jammer and allow normal operation in other beams than the suppressed one. \\

Through the introduction of the Non-3GPP InterWorking Function (N3IWF)\label{N3IWF}, it is also possible to connect to the 5GC using other access networks. \\

As mentioned at the beginning of this document, the 5G standard has been redesigned to make virtualization and cloudification of the core network easier. This also means that most of the technological solutions currently available are actually cloud-based elements of the core network. The introduction of the NRF allows reconfiguration of the network. It also allows the insertion of application functions inside the core network when considered essential for a specific operation. Else, filtered information is provided to third party (less trusted) application functions through the network exposure functions. The general service-based architecture in general uses the same protocol stack (HTTP/2 over TLS) between network functions. At the same time, the 5G network also wants to be able provide low-latency services. The use of a largely centralized core network introduces in some cases too big latencies (just through the geographical distance between a gNB and the data center hosting the core network). The concept of the mobile edge cloud (MEC)\label{MEC} and the use of dedicated UPFs can overcome this limitation by routing some of the data to edge devices that are physically close and thus introduce a minimum processing delay. However, both the cloudification of the core network and the support of network elements in a MEC potentially operated by third parties introduces new security challenges as well. \\

The increased separation between control and user plane decreases the exposure of the 5GC to external network attacks while the implementation of the 5GC through VNFs\label{VNF} might increase the attack surface as they might communicate through an exposed network. \\

In roaming situations, the introduction of the SEPP further hides the network topology of the home network to the visiting network. Indeed, the SEPP serves as a dispatcher of the messages to the appropriate network function without ever having to disclose any implementation details of the network function to the visited network. Furthermore, the use of the SUCI and the modifications in the authentication process now guarantee that only a genuinely authenticated UE can successfully terminate an authentication session. Previously, the MME of the LTE's visited network could fake a successful authentication as it had knowledge of the result of the challenge response exchange. \\

The introduction of network slices not only introduces various QoS levels at radio access network level but also allows separate handling of the slices in dedicated core networks according to their performance and security requirements. In a way, RAN sharing can then also be considered as an extension of the notion of network slices. However, if the network slices are used to implement different security levels, then the right configuration of network slices is essential. 

\subsection{Assets to be protected}
\label{sec4.5}

\subsubsection{User identity and location}
\label{sec4.5.1}

The user’s identity and location shall be protected in privacy. The new concept of transmitting a concealed SUCI instead of the IMSI in an initial registration/attach procedure provides some level of privacy protection. The visiting network is not supposed to be aware of the unconcealed SUPI of the UE until the end of the authentication procedure. At this point in time, the home network effectively has authenticated the serving network to be trusted. Even when the SUPI is transmitted to the AMF of the visiting network, the identity is still not provided to the gNB. However, for some procedures (particularly emergency procedures), the UE will anyway directly communicate its SUPI. \\

Also, temporal, persistent identifiers are still visible during the registration procedures. The core network might also request the device’s IMEI which might allow correlation of a connection with a specific user (particularly if the user connects to both 4G and 5G networks). However, the first NAS message requesting the IMEI is the security mode command sent by the AMF. This message is itself already integrity protected using a key derived from the long-term key stored in the user's home UDM. If the NAS messages are only integrity protected (both on NAS and AS levels), then an attacker could still intercept the information in clear. \\

If an attacker is capable of correlating the 5G-GUTI with the SUPI or PEI/IMEI of a user, it is still possible to track the position of the UE. Indeed, all initial requests in case of the change of the serving cell will still reveal the 5G-GUTI. 

\subsubsection{Service availability}
\label{sec4.5.2}

The impact of denying a device connectivity varies from small annoyance because a phone call cannot be placed to endangering human life if even emergency calls are no longer possible. For machine to machine communications, the systems are expected to be robust in the absence of reliable communications even though the consequences might be anything up to a “graceful” standby of the system.

\subsubsection{Data integrity}
\label{sec4.5.3}

In machine to machine communications, it is important that the data sink can trust the incoming data stream to come from an authentic source. If it is possible to also inject fake data, these pieces of data may not only result in wrong decisions on the receiving end but the level of trust in the authentic data is also decreased. This might then lead either to false alarm type of situations or to a genuine alarm being disregarded by the system. 

\subsubsection{Data confidentiality}
\label{sec4.5.4}

In all communication contexts, the data transmitted over the radio link is the main asset of this link. Depending on the use case, the data may be sensitive and its confidentiality has to be protected. \\

The keys involved in protecting the data both in confidentiality and integrity are secondary assets that naturally also need to be protected. Indeed, leakage of a device’s keys allows an attacker to directly leverage this knowledge to decrypt confidential data and impersonate the device. 

\subsubsection{Network performance}
\label{sec4.5.5}

For safety critical functions, the general availability of the service might be insufficient. Indeed, some devices offer services that rely on low latency being introduced due to the communications channel. Other devices might require a minimum data rate being available at all time to the device in order to perform the proposed service. Thus, devices falling into these categories do not only require the availability of some communications channel but of a communications channel respecting certain boundary conditions. If the network performance is downgraded below a given threshold either in terms of latency or data rate, then for these devices, this situation is equivalent to a complete denial of service condition.
\clearpage
\section{Threats}
\label{sec5}

\subsection{Spoofing}
\label{sec5.1}

\subsubsection{Spoofing of a gNB}
\label{sec5.1.1}

Using software defined radio (SDR)\label{SDR} based solutions, it is quite easy and not very costly to install a fake gNB. Depending on the operator's network architecture (and for example the use of IPsec to protect the N2 interface between the gNB and the AMF), the impact of the fake gNB can vary drastically. If the gNB cannot be integrated into the core network of the attacked network operator, then its impact is limited to misconfiguration of UEs that try to register (or re-register) to the core network through this gNB. Even in this limited case, the previously identified protocol level DoS\label{DoS} attacks on UEs in a 4G network mostly still hold \cite{ref2}. \\

If the fake gNB can be integrated into a legitimate core network, then the fake gNB can still get hold of the $K_{gNB}$ associated to a given UE and it is impossible to identify the gNB as fake by the UE. As the user plane encryption is directly implemented between the gNB and the UE, this also implies knowledge of the key $K_{UPenc}$ and thus the user plane data is directly exposed at gNB level if no end to end encryption is implemented on application level between the UE and the final data sink. However, knowledge of $K_{gNB}$ does not imply knowledge of either $K_{NASint}$ or $K_{NASenc}$. Thus, if the AMF activates NAS level encryption, the gNB cannot generate or intercept NAS messages that request or contain user PII such as the IMEI/PEI. Note that the NAS level identity request message (its reply potentially containing the PEI) is expected to be sent only after successful authentication when the security context has already been established on AMF level. 

\subsubsection{Spoofing of a device}
\label{sec5.1.2}

An attacker might use a fake set of devices to attack the network through the radio interface of a given gNB. However, the long-term key of a UE is stored inside the container of the USIM located in a UICC. This element is subject to security requirements that are intended to guarantee that a key is not easily recoverable even by a motivated and skilled attacker using advanced hardware attacks. However, the initial stages of registering a UE to a network do not require knowledge of the keys. \\

Knowledge of one set of device keys does not allow to multiply their deployment in multiple devices at least not inside the coverage area of the same AMF. Indeed, for each registration procedure, the AMF will generate a new security context based on changed keys received from the UDM. \\

Also, if an attacker can impersonate a genuine sensor, then it is possible to send out manipulated data to the application server in the cloud handling the sensor data. If the sensor data corresponds to an alert, then the generation of false alerts might first of all result in costly deployment of personnel to handle the alert and if repeated over time, the operators might lose trust in the alerts and thus finally choose to disregard even genuine alerts. \\

In case of logistics, the spoofing of a genuine device could be used by an attacker in two circumstances. First, the spoofed device could allow to dissemble the theft of the genuine device or, second, it could be used to inflate the stock of a system.

\subsubsection{Spoofing of a security edge protection proxy}
\label{sec5.1.3}

In case of roaming situations, the home network holding the long-term keys authenticates the requests from a serving network on SEPP level. As specified in 5.9.3.2 of \cite{ref1}, the two SEPPs shall mutually authenticate each other. Indeed, all control plane data between the serving and the home network is expected to pass through the respective SEPPs. Section 13.1 of \cite{ref1} further specifies that TLS (or IPX\label{IPX}) should be used to protect the communication between the SEPPs. The minimum requirements of the supported TLS profiles are further specified in annex E of \cite{ref8}. If an attacker gains knowledge of the credentials of a SEPP, then this can serve as the entry point for the installation of a complete spoofed network impersonating the corresponding PLMN. The spoofed network could communicate with any other network until revocation of the lifted credentials in the other networks. As the initial authentication request from a UE to the home network includes the serving network name and the key derivation procedures also include the serving network name, the spoofed network can however only impersonate this one network. Indeed, the SEPP of the home network is expected to verify the coherence between the used SEPP certificate and the network name received in various NAS procedures. This also implies that the fake network can communicate only with devices that have the given PLMN in the whitelist of networks where roaming is allowed. Also, inside the territory of the home network, this fake network is unlikely to be selected by an UE as the UE will first scan for cells that are part of its home network.

\subsubsection{Spoofing of a security context}
\label{sec5.1.4}
                               
Basin et al. \cite{ref3} mention in their security analysis (end of p. 11 and beginning of p. 12) that the implicit authentication procedure creates potential vulnerabilities in the contexts of on the fly key change and security context switch. However, in our understanding all keys in the system are based on the long-term keys stored inside the UDM of the home network and the UICC of the UE. Replacing a complete security context requires knowledge of these values. Replacing keys in lower layers of the key hierarchy still requires knowledge of keys that are held exclusively in the context of the mobile equipment of the UE or the associated network function inside the 5GC. \\

However, even if the lack of explicit requirements in the 5G standard did not create real vulnerabilities, Basin et al. \cite{ref3} also highlight that security goals are underspecified in the standard. Indeed, in several places of the specifications (e.g. 5.3.4 of \cite{ref1}), the vague notion of a “secure environment” is mentioned without providing a minimum definition of which level of protection constitutes such a secure environment.

\subsection{Tampering}
\label{sec5.2}

\subsubsection{Tampering of a gNB}
\label{sec5.2.1}

A gNB is expected to be updatable through some sort of software update mechanism. If the gNB firmware contains a backdoor (intentionally introduced by an attacker or unintentionally for example in case of a still active debug feature), then this "modified" gNB could potentially expose the user secrets if the backdoor bypasses nominally activated security features (e.g. IPsec). Also, compared to the long-term 5G device keys, the standard does not enforce minimum security levels to be respected by the IPsec implementation. \\

It might also be possible to access memories directly that contain the decrypted data after removal of the IPSec protection.  

\subsubsection{Tampering of a device}
\label{sec5.2.2}

If an attacker can extract the credentials of a UE, then it is impossible to differentiate between a genuine and cloned device. Also, most of the protocol handling is implemented outside the UICC domain of the device. If an attacker can modify the firmware handling the baseband communications of the UE, then it is for example possible to easily generate spurious data transmissions outside the allocated time slots. This might result in a physical/semi-logical denial of service situation for one gNB due to spamming of the random-access channel by a malicious device. 

\subsection{Repudiation}
\label{sec5.3}

\subsubsection{Non-repudiation of a rogue gNB}
\label{sec5.3.1}

The implicit trust in a serving gNB has been reduced compared to the 4G use case. Indeed, the gNB is only in possession of the key used for the protection of the physical channel between the UE and the gNB. Concerning the user identity, outside network functions of the home network only the AMF has access to the SUPI. Therefore, the rogue gNB does not have access to a device’s SUPI. Thus, even if the rogue gNB is accepted by the core network and the attacker only modifies and intercepts the normal exchanges on N1 and N3 levels, it cannot fully track a user directly as the user’s SUPI is still concealed from the gNB. However, as the rogue gNB has been accepted by the system, it is impossible to identify the gNB as rogue and the gNB can still enforce the absence of encryption on RRC and user plane data. Indeed, all devices still have to support null integrity protection and ciphering algorithms. \\

If the rogue gNB acts as a stand-alone gNB that just mimics the characteristics of a normal gNB of the network, then the rogue gNB also has to some extent to implement some of the NAS functions. Given that the standard enforces use of NIA1 or higher as mandatory integrity protection schemes for a number of messages both on RRC and NAS level (except for limited service mode), the rogue gNB can however not impersonate the full network stack of the genuine 5G system as long as the UE fully implements the standard. \\

When a UE switches back from a fake gNB to a genuine one, Annex E of \cite{ref1} describes ways of how the UE measurement reports can be used to some extent in the identification of the rogue gNBs. However, this repudiation of the rogue gNB depends on the measurement configuration sent by the core network to the UEs and on the UEs being able to switch back to the genuine network. 

\subsection{Information disclosure}
\label{sec5.4}

\subsubsection{Data sent by/to device in case of logistics}
\label{sec5.4.1}

In the use case of 5G being used for tracking of stock inside a large organization, the devices probably correspond to a typical mMTC device profile. Indeed, the devices are not required to communicate huge amounts of information and they can be expected to be in a near-stationary location. However, the logistics database containing detailed device ID and associated location information is highly valuable both to the owner of the system as to an attacker. Indeed, depending on the physical protection of the sites where the devices are stocked, knowledge of the position of a particular device might enable an attacker to retrieve it without too much difficulty and with a small probability of detection. Thus, the confidentiality of the data sent by the device is important to be maintained. Given the limited bandwidth allocated to this type of device, it might potentially rely on the link level protection to ensure the confidentiality of its data.
 
\subsubsection{Data interception between entities inside the 5GC}
\label{sec5.4.2}

In chapter 9 of TS 33.501 \cite{ref1}, the use of NDS\label{NDS}/IP is specified between entities inside the core network and between the 5GC and the 5G access network (basically the gNBs). However, it is also noted that there is no need for protection of the communications if the control plane interfaces are trusted. Given that a network operator has the right to decide that they trust their internal network interfaces, it is not guaranteed that any IPsec protection is finally deployed between all entities in the core network or between the core network and the gNBs. If an attacker can bypass the physical security measures, the exchanged data stream might still be directly accessible by the attacker. Depending on the position of the intercept, the exchanged messages can contain various keys that are the result of a UE authentication and registration procedures. \\

Also, if the network is conceived in a way to be highly self-organizing, it will depend on the provisioning policy of the network operator if a new network element can be discovered as not being genuinely part of the network during its installation. \\

It should be noted that all keys derived inside the UE are only used to protect the integrity or confidentiality of data between the UE and AMF or gNB respectively. They are not involved in the protection of data once the data leaves the gNB.

\subsubsection{Identity disclosure}
\label{sec5.4.3}

TS 33.501 (\cite{ref1}) clearly specifies that NAS and RRC signaling data are not required to be protected in confidentiality. This means that a passive eavesdropper might be capable of intercepting the PEI of a UE during the IDENTITY RESPONSE message sent from the UE to the AMF. This intercept also allows to associate the PEI to the current 5G-GUTI of the UE and thus provides tracking information to the attacker. For an attacker, being able to identify the mobile equipment associated to a person is just as valuable as the identification of the identity of the USIM. In case of embedded USIMs (as potentially being the implementation of choice for machine to machine modems due to cost reasons), there is an even stronger correlation between the PEI and the SUPI. \\

The use of the SUCI in the initial registration requests may decrease the level of information that a passive or active eavesdropper can obtain from it. However, it still discloses the identity of the home network. In some contexts, this level of information may be sufficient for an attacker who wants for example track the members of a foreign trade or government delegation. 

\subsubsection{Asymmetric private key disclosure}
\label{sec5.4.4}

The SUPI is concealed using asymmetric cryptography. Currently, this type of cryptographic systems offers a non-brute-forcable level of protection to the asset. However, neither ECC\label{ECC} nor RSA\label{RSA} cryptography are post-quantum resistant. Anyway, if an operator’s private key leaks for any other reason, it is not easily updatable in the UEs meaning that revocation and replacement are difficult to achieve. Knowledge of the private key would allow any eavesdropper to deconceal the SUCI and once again build a SUPI/IMSI catcher for this network operator. 

\subsubsection{Security context disclosure}
\label{sec5.4.5}

The long-term keys and the SQN\label{SQN} are expected to be stored inside the UICC element of the UE. However, practically all other keys are handled outside the UICC and thus less protected against attacks. Though only temporary, this security context can however still have a quite long lifetime. In low-cost modems as used for mMTC, it might be expected that the application firmware and the baseband implementation share the same processor and memory. Then, the security context is potentially exposed through any software vulnerability in the application image.

\subsection{Denial of service}
\label{sec5.5}

\subsubsection{Logical jamming by fake gNB}
\label{sec5.5.1}

If the UE connects to a fake gNB, e.g., because of a better link budget than with the real gNB, the registration procedure can fail and the fake gNB might send a reject cause (e.g. illegal UE, illegal ME, etc.) that results in ROAMING NOT ALLOWED update status. Basically, the UE will not try to register again until the device has been switched off or the SIM\label{SIM} card has been removed and re inserted. In case of machine to machine communications both in stationary and mobile situations, this might result in permanent shutdown of the communications interface if the device is fully compliant with the standard. In case of a stationary device, this registration to a new gNB might be less likely as long as the link budget with the current serving cell remains unchanged.

\subsubsection{Logical or physical jamming of a gNB}
\label{sec5.5.2}

The gNB can only be functional as long as its random-access channel is not jammed. An attacker only needs to jam the gNB during the random-access slots in order to make registration with this gNB impossible. While this type of jamming is not persistent, it impacts all UEs under coverage of this gNB. \\ \\
\noindent
In case of a NSA deployment, the attacker has only to attack the eNB.

\subsubsection{Interruption of the connection to the backhaul}
\label{sec5.5.3}

If an attacker identifies the physical cables connecting a gNB to the core network (or elements of the core networks between each other), then it might be possible to just physically cut the connection. Even if the operator network uses virtualization and cloudification of the network functions to the maximum extent possible, the gNB (or at least its radio part) is still required to be physically present at the cell location to offer its intended service.

\subsubsection{Overloading other network slices}
\label{sec5.5.4}

In case of a RAN sharing situation or of usage of a dedicated network slice for a particular sub system, the availability of the network resources is dependent on the slice orchestration function adjusting the resources to fulfil the service requirements of the devices using the various slices. As indicated in chapter 16 of \cite{ref12}, the resource isolation handling on NG-RAN\label{NG-RAN} level is implementation dependent. If the configuration of the various slices gives precedence to another slice, then it is possible that an attacker connects enough devices to this priority slice and that subsequently the radio resource is depleted for all other slices in the area covered by the saturated NG-RAN. \\

In case of RAN sharing, the physical ownership of the gNBs (and their configuration and control) might be done by the sharing partner. In this case, even an URLLC slice might be subject to this type of attack as the sharing partner might have its own URLLC slice and grant higher priority to its own ultra-reliable slice compared to the sharing partner. Also, in case of RAN sharing, the core network operator has to trust another third-party to not inject tampered data from the gNB side to the 5GC.

\subsection{Elevation of privilege}
\label{sec5.6}

\subsubsection{Physical/logical manipulation of a network component}
\label{sec5.6.1}

If an attacker can modify the firmware inside a gNB, then it is possible to install a parallel communications channel that acts a Man in the Middle (MitM) between the UEs and the network both on control plane and on user plane level. The same reasoning applies to other network functions. If massive cloudification is used for the implementation of the network functions, then the integrity of the cloud servers running the network functions is outside the control of the network provider and another trust relationship is introduced with the cloud service provider. 

\begin{figure}[h]
    \centering
    \includegraphics[width=\textwidth]{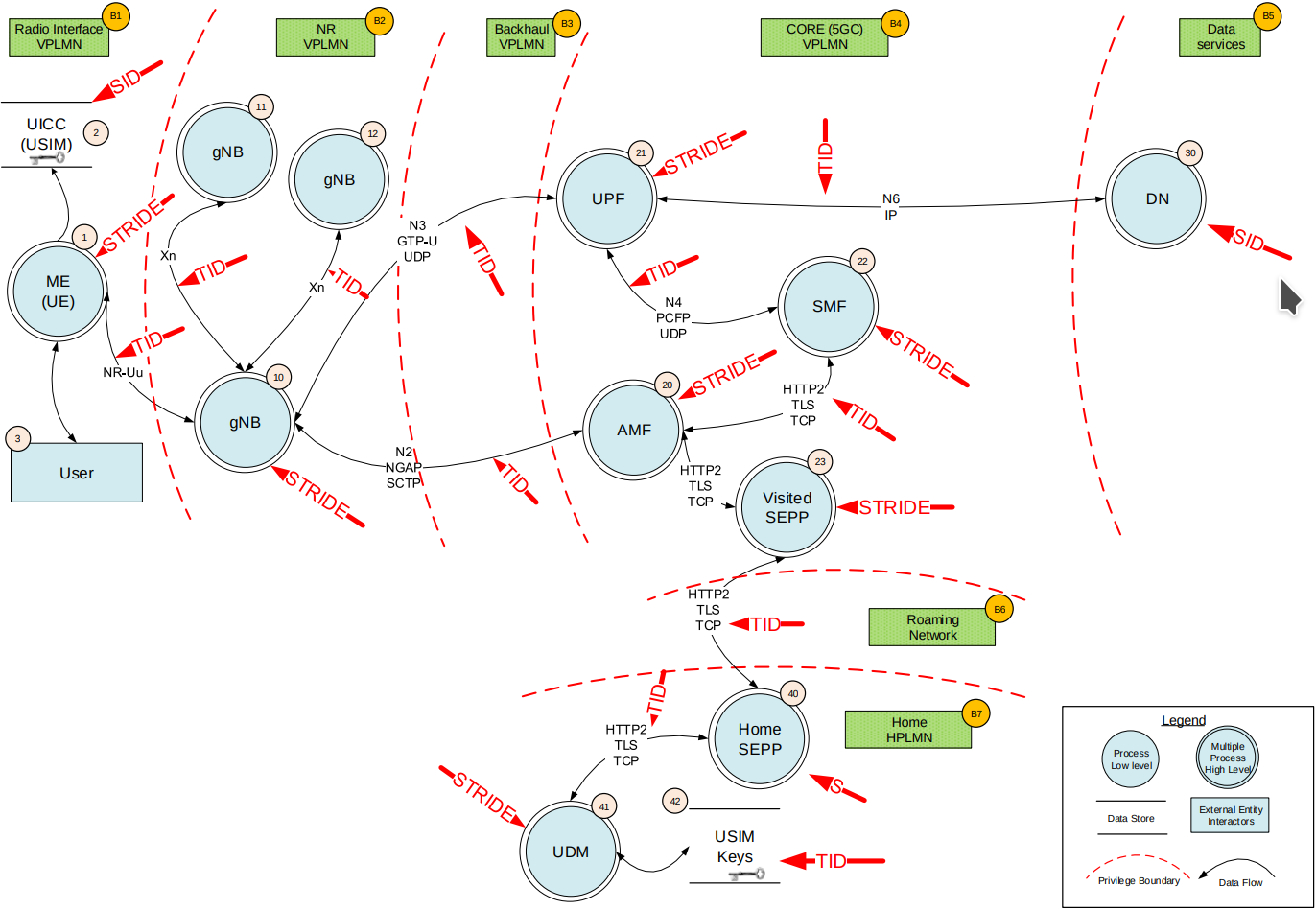}
    \caption{Threats on the various elements in the 5g system}
    \label{fig9}
\end{figure}

\subsection{Threat agents/motivations}
\label{sec5.7}

The threats mentioned above are exploitable by attackers with diverse profiles and distinct motivations.

\clearpage
\renewcommand{\arraystretch}{2}
\begin{xltabular}{\textwidth}{|M{3cm}|M{2cm}|M{2.5cm}|M{4cm}|M{1.5cm}|}
\hline
Threat agent / source                                                                       & Example                                               & Motivations                                                           & Possible actions                                                                                                                                                          & To be considered \\ \hline
\begin{tabular}[c]{@{}c@{}}Non-Malicious\\ -\\ Low capacity\end{tabular}                     & IoT device user                                      & Implementation errors                                                 & Incomplete standard adherence resulting in unintentional DoS.                                                                                                           & Yes                                                          \\ \hline
\multirow{3}{*}{\begin{tabular}[c]{@{}c@{}}Malicious\\ -\\ Medium capacity\end{tabular}}     & Opportunistic Hacker                                 & Fame                                                                  & Hacking as far as possible, publish what has been captured during well-known hacking conferences (i.e. Black Hat…).                                                    & Yes                                                          \\ \cline{2-5} 
                                                                                             & Member of development team, support team, IT admin & Revenge (disgruntled employee, laid off employee)                   & Disrupting one of the 5GC entities.                                                                                                                                      & Yes                                                          \\ \cline{2-5} 
                                                                                             & Criminal                                              & Financial benefit                                                     & DoS attacks and associated blackmail.                                                                                                                                    & Yes                                                          \\ \hline
\multirow{2}{*}{\begin{tabular}[c]{@{}c@{}}Malicious \\-\\ High capacity\end{tabular}}       & Criminal organizations                                & Financial benefit, untargeted                                        & Ransoming, DoS attacks, tracking high value victims, detecting law enforcement officers.                                                                               & Yes                                                          \\ \cline{2-5} 
                                                                                             & Terrorist organizations                               & Multiplying impact of terrorist actions                              & Disrupting critical infrastructure, finding targets for attacks                                                                                                         & Yes                                                          \\ \hline
\multirow{2}{*}{\begin{tabular}[c]{@{}c@{}}Non-malicious\\ -\\ Medium Capacity\end{tabular}} & Security Researcher                                   & Poor security or operational errors                                  & Analyze security of system as far as possible, publish in journals or conferences to show their expertise.                                                             & Yes                                                          \\ \cline{2-5} 
                                                                                             & Member of development team, support team, IT admin & Poor security, operational errors                                    & Unintentional information leakage, loss of availability, mistakes with installed software or hardware.                                                                 & Yes                                                          \\ \hline
\multirow{2}{*}{\begin{tabular}[c]{@{}c@{}}Non-malicious\\ -\\ High capacity\end{tabular}}   & Competitors                                           & Business opportunity                                                  & Exploit known or published information. Could provide support to activists, hackers or security researchers to publish weaknesses on the products inside the system. & Yes                                                          \\ \cline{2-5} 
                                                                                             & Hacktivists / Politics                               & Highlight risks with this type of solution                           & Make bad publicity on the 5G system and relay weaknesses published to discredit the system or the telco                                                                & Yes                                                          \\ \hline
\begin{tabular}[c]{@{}c@{}}Malicious\\ -\\ Unlimited \\ capacity\end{tabular}                  & Foreign government agencies                          & Intelligence gathering, damaging reputation of the targeted country & Intercepting sensitive data, tracking VIP targets, disturbing critical communications                                                                                   & Yes                                                          \\ \hline
\caption{Threat agents and their motivations.}
\label{tab6}
\end{xltabular}

\subsection{Threat vectors}
\label{sec5.8}

The table below provides an estimation of the effort and expertise needed for some attack scenarios on various parts of the 5G system. These attack vectors are covering both remote and local attacks. Some of the local attacks could be a first step to find new remote attack vectors that are then potentially scalable to a bigger number of devices or that might provide the means of attacking more central network components.

\begin{xltabular}{\textwidth}{|M{3cm}|M{1.1cm}|M{5.4cm}|M{1.5cm}|M{1.5cm}|}
\hline
Remote or Physical attacks                                                                                                                                             & Budget set up                                         & Example of attack scenarios                                                                                                                                                                                                                                                                                                                                                                                                   & Expertise                                                      & Effort MDs                                                                                                                             \\ \hline
Remote message interception :\newline Interception of 5G-NR messages                                                                                                             & 1000 \$                                                & - Attacker uses a SDR to intercept all exchanges between the UEs and a given gNB                                                                                                                                                                                                                                                                                                                                             & Knowledge of common 5G radio implementation                    & - Some hours to set up \newline - Needs to be listening when the exchange occurs                    \\ \hline
Fake gNB :\newline Broadcasting of manipulated signaling messages                                                                                                                & 5000 \$                                                & - Attacker implements a rogue gNB with limited AMF implementation to force registration of UEs to this gNB and either intercept temporary UE identities or perform denial of service attacks on UEs trying to register.                                                                                                                                                 & Knowledge of 5G registration procedures and NAS protocol       & - Some hours to set up \newline - Needs to be the gNB of choice for a UE                              \\ \hline
Physical jammer Jamming the receiver of a gNB                                                                                                                          & 5000 \$                                                & - Attacker uses a directive antenna to directly point at the antenna element of a gNB and transmits coherent or incoherent signal in the 5G band of the gNB with significantly higher effective power than real UEs.                                                                                                                                                 &                                                                & - Need to remain in vicinity of the gNB                                                                                                \\ \hline
Physical manipulations Observations or simple manipulations on insufficiently protected network link                                                                   & 100 - 500 \$                                          & - Physical observations and manipulations on network interfaces or the cables between elements of the 5G network (particularly gNB and backbone) \newline - Network analyzer to dissect network traffic.                                                                                                                                                                             & Very limited electronic skills, some physical intrusion skills & - Some hours to set up                                                                                                                 \\ \hline
Reverse engineering on SW\label{SW} application using tools (IDA pro…)                                                                                                           & 5 K                                                   & - Software reverse activities to identify flaws in software when binary has been extracted from internal / external memories then exploit the vulnerabilities to, e.g., push malicious FW to the entity running the binary or extract secrets accessible by the SW.                                                                                                  & Hackers 500 \$ per day                                          & 30 MD for software reverse                                                                                                             \\ \hline
Non-invasive fault injection \newline- Electrical Glitch with low cost equipement \newline- Electrical Glitch with lab equipement          & 1 K \newline 10 K & - VCC / GND / CLK electrical Glitch on MCU to inject fault into software and bypass critical checks / operations done by software without countermeasures against fault attacks.\newline- Electrical Glitch allows an attacker to dump secret from RAM\label{RAM} (e.g. Keys) or dump internal memories.\newline- Fault injection might allow bypass of authenticity checks during FW update installation. &                                                                & - 25 MD for set-up \& attack scenario \newline - Few minutes / hours to repeat the attack \\ \hline
Non-invasive: Side Channel Analysis (VCC / EM)                                                                                                                         & 30 K                                                  & - SCA\label{SCA} attacks (DPA, DEMA, template…) to retrieve secret keys during execution of crypto algorithms (ex AES, ECDH)                                                                                                                                                                                                                                                               & Lab Expertise 1 500 \$ per day                                  & - 25 MD for set-up \& attack scenario \newline - Potentially only few hours/ days to repeat the attack \\ \hline
Semi-invasive \newline - Fault injection: Laser or other means to extract private keys of a NF (objective of choice SEPP) & 50 K                                                  & - Fault attacks to change protection state and dump internal memories \newline - Multi-fault attacks to retrieve secret keys during crypto-algorithm execution (ex AES, RSA, ECDH)                                                                                                                                                                                                       &                                                                & - 25 MD or less for set up \& attack scenario \newline - Few hours to repeat the attacks                 \\ \hline
Invasive attacks on hardware Physical attacks on MCU\label{MCU} or memory with SEM\label{SEM}, FIB\label{FIB}…                                        & 1 M                                                   & - Physical attacks to reverse internal memories of UICC and get firmware and data.                                                                                                                                                                                                                                                                                                                                          &                                                                & Up to 200 MD or more                                                                                                                   \\ \hline
\caption{Threat vectors and their associated complexity.}
\label{tab7}
\end{xltabular}

One of the central assets are the keys held inside the UICC on device side and the UDM on network side. Thus, if an attacker can extract the database of device keys from a network’s HSM\label{HSM} inside the UDM, then the attacker has full control of the communications with all devices handled inside this database. However, getting into possession of this database is expected to first require physical access to the storage devices of the UDM and, second, potentially HW\label{HW} attacks to bypass the physical security elements of an HSM. If the theft of a HSM is detected in time, it might be possible to update the keys of the potentially compromised devices before the attacker had time to extract the keys from the hardware element. \\ \\
\noindent
Concerning extracting the keys from the UICC on UE side, it is considered to be potentially more easily achieved. However, this sort of attack would only provide access to the keys of a single UE. 
\clearpage

\section{Threat Scenarios}
\label{sec6}

\begin{figure}[h]
    \centering
    \includegraphics[width=\textwidth]{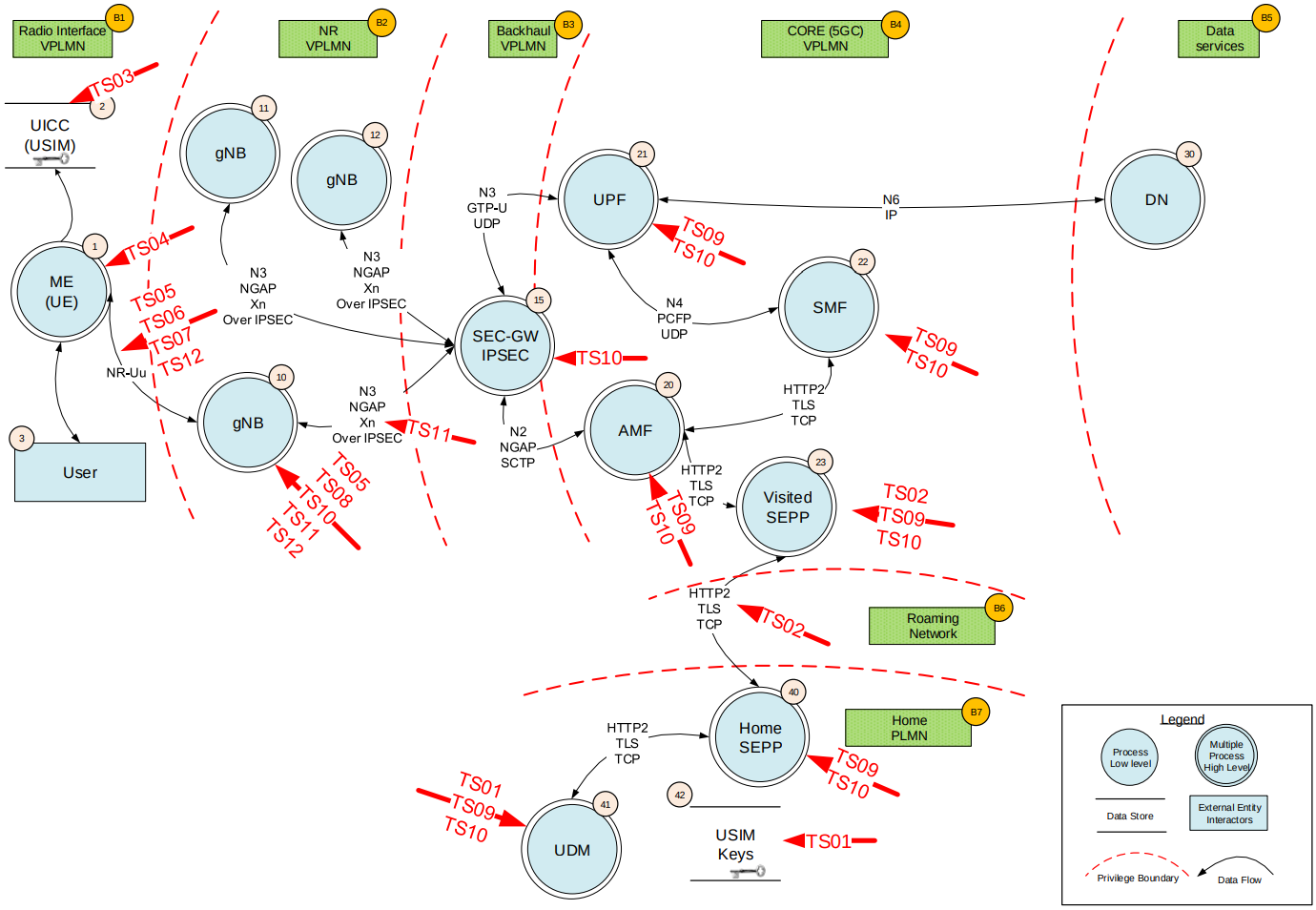}
    \caption{Location of the threat scenarios in the DFD} 
    \label{fig_threat_data_flow}
\end{figure}

The table below lists a summary of potential threat scenarios. The various scenarios are then explained in more detailed in the consecutive sections of this chapter. Figure \ref{fig_threat_data_flow} provides a view on the position of the threat scenarios in the data flow diagram.

\begin{xltabular}[]{\textwidth}{|M{1.5cm}|M{1.5cm}|M{3cm}|M{3cm}|M{1.5cm}|M{1.5cm}|}
\hline
STRIDE & Assets                                                                       & Threat scenarios                                                                                                                                                                                                                                                                                                                          & Context and potential security controls                                                                                                                                                                                                                                                                       & Impacts                       & Likelihood and ID           \\ \hline
STRIDE & Long term keys of UEs in a given network                                  & A disgruntled employee with access to the database of all device key makes a copy of the keys and sells them to a criminal organization.                                                                                                                                                                                             & The UDM manages all keys used inside the network. Security control: strict access control and use of HSM to protect the keys, update mechanism of keys stored in the operator’s UICCs                                                                                                                   & Critical                      & Unlikely TS\_01             \\ \hline
SRIE   & Private key used by a SEPP to authenticate to other networks                 & Extracting the private key used by the SEPP to authenticate itself to other networks through side channel attacks, an attacker then impersonates the SEPP and spoofs a roaming network with the identity corresponding to the extracted key.                                                                                          & Networks authenticate each other through the authentication of the respective SEPPs Security control: Data analytics and detection of anormal activity level of a serving network followed by potential key revocation                                                                                 & Very High                     & Unlikely TS\_02             \\ \hline
SRIDE & Device keys Data transmitted/ received by a device Device service continuity & Key extraction through hardware attacks on the UICC element \newline - HW expert first xtracts the keys from the UICC of a valid device. Keys used to create clones and attack the network or if attack is on-invasive/ destructive also to spy on communications of the legitimate user                                                    & Difficulty depends on robustness of UICC                                                                                                                                                                                                                                                                     & High (limited to one device) & Unlikely TS\_03             \\ \hline
SRIDE & Device security context Data confidentiality and integrity                   & A malware on the ME with sufficient privilege dumps the current security context of a device. The dumped keys can then be used to impersonate the device to the network and to decrypt all previous communications of the device                                                                                                  & Keys derived in the context of a registration procedure are held outside the UICC in the context of the MESecurity control: Regular renewal of the device security context by the network                                                                                                               & High (limited to one device) & Probable TS\_04             \\ \hline
D      & Service availability                                                         & Physical or Logical jamming of devices through fake gNB                                                                                                                                                                                                                                                                                   & - Impact per jammer limited to its coverage. \newline - Except for protocol based jamming during the attach procedure of a device, duration of impact only as long as the jammer is active Security control: Blacklist of fake gNB broadcast in nominal network                                                & Moderate                      & Probable TS\_05             \\ \hline
I      & Location tracking                                                            & Partial SUCI and PEI catcher through interception of radio link                                                                                                                                                                                                                                                                           & Security control: Encryption of signaling messages both on radio and NAS level to protect PEI                                                                                                                                                                                                                & Moderate                      & Very Probable TS\_06       \\ \hline
D      & Service availability                                                         & Physical or Logical jamming of gNB                                                                                                                                                                                                                                                                                                        & - Impact per jammer limited to one gNB. \newline - Impact only as long as the jammer is active Security control: BFN to eliminate the jammer’s radio signal                                                                                                                                                       & Moderate                      & Probable TS\_07             \\ \hline
TRIDE & Service availability, Data confidentiality and integrity, device location    & Exploit of software vulnerability in a gNB (or malicious firmware update) to install backdoors to data buffers and extract signaling information in clear or might result in attacker managed DoS                                                                                                                                      & - Tampered gNB might share handled data \newline - Might provide access to gNB level keys- Vulnerability might be built in unintentionally or by malicious supplier and actions triggered through radio interface Security control: External audit of gNB code and secure coding rules Authentication of firmware & High to Catastrophic          & Probable TS\_08             \\ \hline
TRID   & Service availability, Data confidentiality and integrity,                    & Exploit of software vulnerability in a network function (or malicious firmware update) can lead to misconfiguration of UEs, data leakage and bypass of security controls ; in a virtualized network function this can include data leakage through side channel attacks between virtual machines using the same physical resources & Tampered NF (e.g. AMF) might disclose current security context of a device or not implement all optional security features                                                                                                                                                                                & Very High                     & Unlikely TS\_09             \\ \hline
TI     & Device data                                                                  & Extraction of keys used to establish IPSec connection from link node memory - If a link node (gNB, AMF, UPF, etc.) uses SW implementation of IPSec, keys might be exposed through heartbleed style of attacks \newline - In gNB, they might not be stored in secure storage and extracted through local physical access                   & - SW vulnerabilities. \newline - SW implementation of cryptographic suites. Security control: Use of robust hardware module for handling of root keys used for secure channel establishment                                                                                                                       & High                          & Very probable TS\_10        \\ \hline
D      & Service availability and network performance                                & Stealing or modifying the physical configuration of a gNB \newline - Disrupting access to the backhaul \newline - Removal of gNB or its antennas in insufficiently secured physical location                                                                                                                                                          & Mitigations: \newline - Physical security for gNB access. \newline - Overlap in the cell coverage.                                                                                                                                                                                                                           & Moderate                      & Very probable TS\_11        \\ \hline
D      & Network performance                                                          & Overloading traffic in high priority slice in respect to "lesser" priority slice (or slice associated to another PLMN in RAN sharing case)                                                                                                                                                                                             & Mitigations: \newline - Proper implementation of service level agreements and resource management function in gNBs                                                                                                                                                                                                   & Moderate                      & Unlikely to Probable TS\_12 \\ \hline
\caption{List of potential threat scenarios.}
\label{tab8}
\end{xltabular}

\subsection{TS\_01 Operator UDM database theft}

The keys contained in the UDM database are also stored in the UICC elements of the UEs. Having dropped this database, an attacker can fully impersonate the network. Also, it is not very easy to update the long-term keys in the UICCs thus it is very costly to mitigate this sort of attack. For massive IoT (MIoT)\label{MIoT} deployments or for elements embedded in some HW modem that is not easily accessible inside a vehicle ECU\label{ECU}, it is difficult to access the UICC and to physically replace it if this is considered more feasible than a root key update. \\

The mitigation to this attack is the physical access control to the UDM that is expected to be very strict. Also, using an HSM to protect the keys probably means that the attacker still has to do time consuming attacks once in possession of the HSM to extract the data. This time window might be sufficient for the operator to be aware of the loss of the device and to deploy new keys in the UICCs of their network. \\ \\
\noindent
\textit{Threat agents:} Malicious employee (or employee under duress due to criminal blackmail) with access to the UDM storage. Given the amount of confidential information being disclosed through one attack, the motivation for a criminal organization or hostile nation can be considered high. 

\subsection{TS\_02 Impersonation of a roaming partner through knowledge of the authentication key of the SEPP}

Being able to impersonate a roaming partner, an attacker can build a complete network with gNBs under their and use this parallel network to intercept and tamper data received and transmitted by a UE. The rogue network would also be in possession of the SUPIs of UEs registering through it. \\

To impersonate the roaming partner, the attacker would in priority have to obtain the private key of the roaming partner’s SEPP used to authenticate it to the attacked network. This might be possible using logical exploits of the authentication is using software processes or non-invasive attacks to extract the key for example trough side-channel attacks. Then, the attacker would have to take down the SEPP of the impersonated network in order to be able to replace it in regard to the \say{home} network. If the attack is perpetrated by a nation-state, then the attacker can possibly just request the credentials from the impersonated network. As a final step, the attacker has to deploy infrastructure for the rogue 5G network. \\

However, if a UE captures cells both from its home network and this rogue network, it will in priority connect through the cells of its home network. Thus, this rogue network would potentially only capture UEs that are in a real roaming situation or the attacker would still have to physically disrupt the original home network, e.g. by destroying the antennas of the real network under attack. \\ \\
\noindent
In summary, the cost of this sort of attack risks of being rather high for an attack having only this impact. \\ \\
\noindent
\textit{Threat agents:} Criminal organizations, foreign government agency.

\subsection{TS\_03 Device long term key extraction through HW attacks on the UICC element}

The UICC contains the keys used at the root of the key derivation and agreement process between the UE and the network. If an attacker can extract the key material from a legitimate UICC, the key material can be used for various purposes. First, the attacker can generate clones of the device. However, the network should only authorize one active security context at any given time and thus the cloned (and legitimate) devices cannot function parallelly. \\

If the keys have been extracted without the attack being detected by the device owner then, their knowledge allows permanent eavesdropping on the exchanged data and injection of fake data. Tampering detection is naturally more difficult to achieve if the UICC is used in a M2M context. Another means of allowing an attacker enough time to extract the secrets would be their extraction before the initial use of the UICC in a UE. \\ \\
\noindent
Note that this attack only exposes the secrets on a UICC per UICC base. \\ \\
\noindent
\textit{Threat agents:} Security researchers to check the robustness of products and test their technical capabilities, criminal organizations and foreign government agencies. 

\subsection{TS\_04 Non-permanent key extraction from mobile equipment} 

As shown in Figure 6.2.1-1 of \cite{ref1}, most of the keys inside the UE are handled inside the ME and not the USIM. While the security requirements are very well specified for the USIM by GSMA\label{GSMA} and in 5.2.4 of \cite{ref1}, the requirements are less clear for the ME. Even though the baseband and application space inside normal UEs are quite often separate sub-systems, it is possible to imagine that the application and baseband processing might be handled in the same processor particularly for low cost components such as devices used in mMTC for example in the context of logistics (real time package tracking, stock management, etc.) and battery powered sensors in smart cities. \\

Given that there is no explicit requirement on how to protect the keys handled by the ME in the context of its current security context, it cannot be excluded that a malicious application running inside the ME does not have access to this security context. Knowledge of the current security context allows an attacker to eavesdrop and inject messages nominally from the UE to the network. In case of a sensor device, sending fake data might for example result in a false alarm and result in costly dispatching of a maintenance team. \\

Unless the network triggers the renewal of the security context, these keys will remain valid. For a stationary IoT device, the network might want to limit the amount of exchanged data (including signaling data) with the device to a minimum and therefore refrain from renewing the security context within relatively short intervals. This means that the extracted security context can have a nearly permanent validity. \\

Depending on the security mechanisms used by the ME protecting against the installation of malware, this attack can be much easier to perform than TS\_03 with a nearly comparable result. Also, in the context of IoT, it might be more attractive for hackers to attack the new IoT device compared to the “boring” well-known USIM. This adds motivation to this sort of threat agent. \\

Even if more complex ME architectures are used, it is expected that the extraction of a security context from the ME is much less costly than the extraction of secrets from the UICC. \\ \\
\noindent
The extraction of the security context can however only be achieved once the device is operational. \\ \\
\noindent
\textit{Threat agents:} Opportunistic hacker, criminal, security researcher. 

\subsection{TS\_05 Physical or logical jamming of devices}

Physical jamming of devices will only affect the UE as long as the jammer is active. Depending on the covered frequency bands and on the beamforming capabilities of the device, the device might even be capable of blocking the angle of arrival of the jammer. In case of a logical jammer however, attacks equivalent to known 4G attacks (cf. \cite{ref2}) are once again possible and their impact will persist until the device has undergone a power cycle. Indeed, if a UE tries to switch to this rogue gNB following the cell selection and reselection mechanism described in \cite{ref6}, then the rogue gNB can trigger a new registration procedure followed by transmitting an unprotected REGISTRATION\_REJECT NAS message. As stated in 4.4.4.2 of \cite{ref11}, this message has indeed to be processed before a valid security context is established between the UE and the network. \\

In case of stationary IoT devices (e.g. sensor devices, smart meters, etc.), the cell reselection criteria might be difficult to achieve by the rogue gNB as long as the current cell on which the device is camped remains powerful enough. For highly mobile IoT devices as in the modem ECU of an autonomous vehicle relying on V2X communications to interact with its environment, the device is however constantly changing cells. In this case, the rogue gNB just needs to provide a slightly better signal than another candidate cell of the attacked network to be the cell of choice in the cell selection process. Given that some rejection causes require the device to either follow a power cycle or to have its USIM being reinserted to be active again, this logical jamming can indeed have a near permanent effect on mobile IoT devices. For example, a drone being controlled through 5G would naturally either have to disregard the 5G specifications or go into a safe return mode as there would be no means of a human manually triggering a power cycle while flying. \\ \\
\noindent
The cost of the rogue gNB can actually be estimated to be lower than a high-end physical jammer. \\ \\
\noindent
\textit{Threat agents:} Criminal and terrorist organizations.

\subsection{TS\_06 Location tracking through interception of normal radio link}

Depending on the choice of the network operator, signaling messages can only be integrity protected. Even though the SUPI will still only be transmitted in its concealed form, an attacker can still gather the same amount of information through the home network identifier transmitted in the context of the authentication procedure and the PEI transmitted inside the SECURITY MODE COMPLETE message. If the network operator chooses to use encryption for signaling messages, then an attacker can only capture the SUCI and the associated home network identifier. If the home network is one of the swiss networks, then this is not really any differentiator for the attacker. If the home network is more uniquely identifiable (e.g. visit of a foreign delegation attending the WEF), tracking all phones associated to a given home network might be of enough interest to an attacker. \\

If the attacker disposes of a network of (potentially low-cost) radio sensors with sufficient density, then it might be possible to track the location of a given set of UEs at all time. With the knowledge of a vector of positions of the targeted user, big data analysis might actually allow to match the known position vector to a 5G-GUTI without knowledge of either SUPI or PEI and to complete the location pattern of the targeted UE outside the previously known locations. More importantly, with the knowledge of the location of the target at the beginning of the tracking session, it might be possible to track the target without physically following it after this initial matching phase. \\ \\
\noindent
\textit{Threat agents:} In case of the absence of encryption of signaling data, the location tracking might be of interest to criminals, terrorist organizations or foreign government agencies. If only the home network identifier can be intercepted, foreign government agencies might still remain motivated to implement this type of attack. If the tracking is based on a sensor network, then it can be assumed that only foreign government agencies have the resources to install this type of network. 

\subsection{TS\_07 Jamming of a gNB}

Contrary to the logical jamming of UEs that will impact the operation of the UE beyond the presence of the jammer, the effect of the jammer on a gNB will disappear as soon as the jammer is no longer active. From a protocol point of view, it should however be quite easy to obtain a modified rogue UE that jams the random-access channels of a gNB. In this case, the jammer would indirectly deny a new UE to request access to the cell using random access procedures and the gNB would be severely impacted in its operations and network performance for this cell would decrease drastically. \\

If the gNB detects the presence of this logical jammer and is capable of locating its position, then the gNB might configure its BFN in a way to suppress the direction of arrival of the jammer signal. This suppression capability will however be dependent on the size of its antenna array (and indirectly on cell center frequency). \\ \\
\noindent
This attack would only impact a single gNB. \\ \\
\noindent
\textit{Threat agents:} Criminals.

\subsection{TS\_08 Malware or SW vulnerabilities on gNB}

The software stacks inside a gNB and network functions of the 5G core are complex. Also, the manufacturers of the equipment might not be willing to share the code even with the network operators as the scheduling function might contain highly proprietary optimizations. The SW of a gNB is expected to be updatable. \\

If vulnerabilities are present either because of backdoors mandated by the government of the equipment manufacturer, due to coding errors or after replacement of the original firmware by a malicious firmware, then the modified gNB would expose a behaviour that is no longer in accordance with the specifications. This could lead to the gNB being taken down or providing access to confidential data to the attacker. If the vulnerability is already present in the official firmware, it might be exploitable through the radio network. In this case, all gNBs with the same vulnerability would be at risk and the result could indeed be catastrophic for the infrastructure of a network operator or even the country. \\

The modified gNB could also be used as entry point to attack core network functions through the existing network link between the gNB and the core network (particularly the UPF and the AMF). However, the feasibility of this attack depends on the absence of any load balancer in front of the 5GC. \\

Thanks to virtualization concepts, the gNB can be implemented in a split architecture and the non-time critical sections of the gNB-CU can also be located in the cloud. This cloud instance can handle more than one physical RAN. In this case, a successful attack on the cloud instance (through for example physical access to the data center hosting the cloud VM\label{VM}) directly impacts more than a single physical gNB instance. \\ \\
\noindent
\textit{Threat agents:} Disgruntled member of development team for malicious inclusion of backdoor in the firmware code base, member of the development team unintentionally inserting exploitable vulnerability into the firmware, security researcher analyzing the firmware and detecting a vulnerability, government agency mandating inclusion of a backdoor into code provided to foreign operators that the mandating government agency can activate at will. 

\subsection{TS\_09 Malware or SW vulnerability in 5GC network function}

Comparable to the gNB, an attacker might be able to exploit a vulnerability in a network function such as the AMF to get access to the parts of the security context handled by this network function. Given the key derivation schemes applied in 5G, knowledge of lower level keys (e.g. KgNB) does not provide knowledge of higher level keys (e.g. KAMF). However, this reasoning does not apply in the other direction. A misconfigured SMF could also instruct the gNB to configure the data bearers as not being confidentiality protected. \\

As the network functions do not require to be distributed to cover the territory of the operator, they can be located in physically secured locations. This makes a local attack on the network functions less likely. \\

If virtualization is used, they can also be operated from the cloud and thus be physically hosted in data centers of cloud service providers. However, this dependence on the cloud service provider introduces a new trust relationship that is not handled in the specifications. The data centers might even be located in a different country with differing legal requirements in terms of data protection. Furthermore, the virtual network function might run on a cloud instance that also executes an attacker's virtual machine and that might be able to access the network function's context through micro-architectural attacks or vulnerabilities in the hypervisor managing the various virtual machines. \\ \\
\noindent
\textit{Threat agents:} Opportunistic hacker if the control interface of the network function is exposed on the public internet, criminal organizations for blackmailing the network operators, government agencies for espionage and control of foreign infrastructure. 

\subsection{TS\_10 Lifting of keys used for link protection between network functions}

In the best case, the network operator does not rely on the physical security of their network to protect the data in transit between different network functions. However, depending on the IPSec implementation, it might be possible to extract the keys used for securing the link through a zero day exploit on the software running inside the network function or through side channel leakage to other functions being executed on the same hardware. If the network equipment is accessible, an attacker might also choose to use physical attacks to extract the keys used for the protection of the network link. \\ \\
\noindent
\textit{Threat agents:} Criminals, hackers, security researchers.

\subsection{TS\_11 Theft or physical misconfiguration of a gNB}

Depending on the type of cell implemented by a gNB, access to its antenna might be more or less difficult. Accessing the connection of the gNB to the backbone is expected to be even more difficult to protect. Given the skepticism related to 5G radio transmissions in parts of the population, it is possible to imagine that a small community of hacktivists decides to disregard planning or court decisions and actively removes or destroys the antennas of 5G base stations whenever easily accessible. \\ \\
\noindent
\textit{Threat agents:} Hacktivists.

\subsection{TS\_12 Exploit of bad resource management in slice resource allocation}

In case of RAN sharing and in a lesser extent also in slice management for a single operator, the risk exists that the radio resource management might fail to allocate sufficient resources to high priority slices (e.g. URLLC) when other slices face a high load. In case of RAN sharing, the primary owner of the radio resource might privilege its own radio resource requirements and no longer guarantee sufficient bandwidth to the sharing PLMN in case of network overload. Apart from the generic network overload aspect, this attack will however heavily depend on implementation choices made by the network operator. \\ \\
\noindent
\textit{Threat agents:} Criminals, terrorists.

\clearpage
\newpage
\section{Risks Matrix}
\label{sec7}

\begin{figure}[H]
    \centering
    \includegraphics[width=\textwidth]{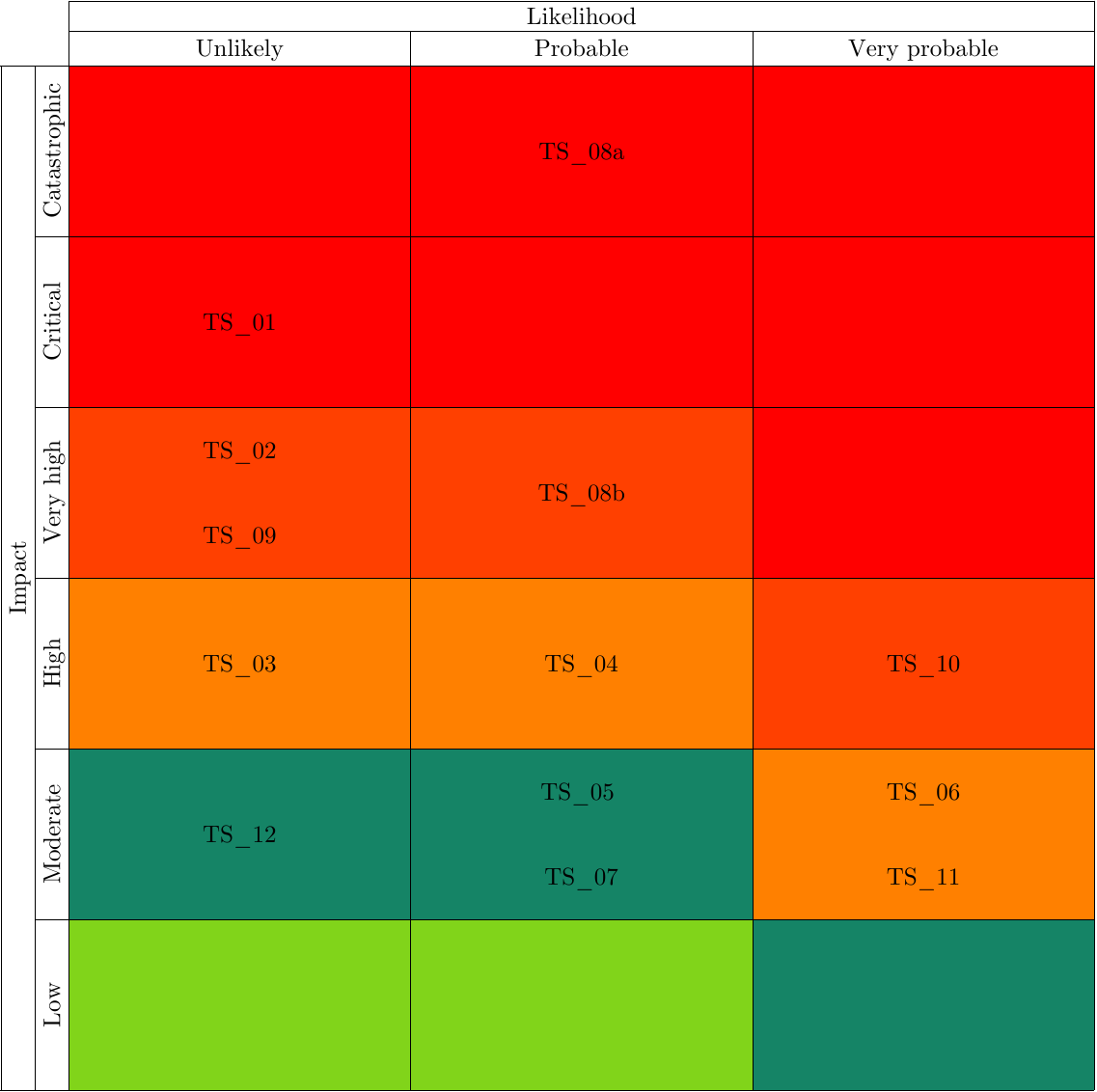}
\end{figure}
\clearpage
\section{Mitigations and Security Controls}
\label{sec8}

Several of the threat scenarios are only possible due to under-specification of the 5G standard. Indeed, if an operator implements all optional security and follows the recommendations inside the specifications, then some of the scenarios are actually impossible to exploit. \\

Other attack scenarios rely on an insufficient level of protection of security features. Indeed, security is not achieved by merely activating a feature but activating a feature in a robust manner that withstands attacks against its bypass or deactivation. \\

Concerning TS\_01 (lifting of the key database of the subscribers), if it is possible to update the keys in the UICCs used by the network operator and if the used HSM is sufficiently robust, then it might be possible to mitigate the attack before the attacker has been able to extract the keys of the lifted database. However, the robustness of the protection mechanism of the database in the UDM will be highly be highly dependent on its logical and hardware implementation. It might even be possible that the operator is dependent on the physical security of their cloud service provider. \\

In roaming situations and to some extent also in case of external application functions, the operator extents the trust boundary to a third party (TS\_02). Without data analytics discovering anormal behavior of this third party, the operator then depends on the third party applying the same level of security controls as themselves. It is however unlikely that a Swiss network operator will audit the security of all their roaming partners. Data analytics might allow the discovery of anormal amounts of data exchanges. The operator might then check with the roaming partner if number of exchanges correspond to the expected behavior on the other side and result in revocation of the roaming partner's certificate in case of a mismatch. \\

Extracting the keys of a single subscriber through the attack of the associated UICC (TS\_03) might be made more difficult by using hardware elements with additional countermeasures against both passive and active attacks. External certification of the UICC might provide an increased level of confidence in its robustness. \\

Attacks that are based on potential vulnerabilities or non-compliances inside the ME of the user equipment (TS\_04 and TS\_07) can only be mitigated by the network operator inside the core network. Indeed, the network operator only has control over the UICC inside the terminal. Knowing that the trust in the security of the ME is limited, the network operator should force a renewal of the security context on a regular basis (TS\_04) in order to limit the duration of a security breach the and be able to suppress some directions of arrival to filter out logical and physical jammer signals (TS\_07). \\

In the current version of the specifications, a compliant device has no means of mitigating logical jamming attacks of some REGISTER REJECT causes sent by the rogue network (TS\_05). Indeed, this message can be sent before the establishment of a security context and the network currently has no means of authenticating itself before the security context has been configured between the network and the device. A potential mitigation for this situation could be as follows: All currently global reject causes should be limited to a single network. The network would identify itself by broadcasting a network pre-security context authentication public key (e.g. in one of the system information blocks) and signing the reject message using the associated private key. Thus, an attacker without knowledge of the real network private key cannot fully impersonate this network. A rogue gNB could naturally broadcast its own public key and reject the registration of any UE. However, the UE would still be authorized to try to re-register to another network broadcasting a different public key. Note that currently the impact of a fake gNB is potentially much higher for an IoT device (and particularly a moving IoT device) than for a normal mobile phone. If the real network is made aware of the presence of a jammer in one of its cells, it can also blacklist this rogue gNB in the system information broadcast by the surrounding legitimate gNBs. \\

The disclosure of the PEI described in TS\_06 is only possible if the network operator chooses not to apply NAS and radio level encryption for control plane messages. \\

The exploit of vulnerabilities allowing extraction of key material or tampering with the firmware in the gNB or other network elements (TS\_08 to TS\_10) depends on the robustness of the authentication functions of these functions at boot time and the presence of vulnerabilities inside the software. As these functions are essential or the correct operation of the network, the network operator should be aware of their importance and implement procedures that allow to increase trust in the correct and robust implementation of these functions. This applies both in regard to the equipment manufacturer as well as other service providers (e.g. cloud operators). The operator should also evaluate the design features used to protect the authenticity executed functions and confidentiality of secrets. \\

Concerning the threat scenario TS\_11, increasing the acceptance of 5G systems by open discussions with the public should at least reduce the risk of destruction of base stations by hacktivists. To avoid network disruption by criminals or terrorists, the physical security of the access to the base stations and redundancy in cell coverage are the only means to maintain network operations at all times and in all places. \\

For TS\_12, appropriate resource management between slices taking into account their criticity and general QoS requirements should allow to mitigate this threat.

\clearpage
\section{Conclusion}

5G networks are still exposed to a number of threats previously identified in the context of 4G implementations. This naturally remains even more true in non-standalone deployments where the network is 5G in name only (or to be more precise only 5G for some aspects of the radio channels). Some of these threats have an increased impact due to new use cases that are expected to be increasingly significant in 5G networks such as massive M2M communications with battery powered devices requiring low data rates and M2M communications with high reliability and low latency constraints. The main challenge for these device classes is that interaction with a human user is limited and remediation of attacks cannot depend on any direct actions of the user. Another challenge is that due to the specific performance requirements of these devices, the network operator might be tempted to not use all possible security controls (e.g. user plane encryption and integrity protection) for communications of these device classes. Finally, the virtualization concepts create new challenges to the operators as they potentially create new trust relationships between the operator and third parties such as cloud service providers. 
\section{Future Work}

Outside the context of UEs in limited service state, exchanges with the gNB on RRC level and with the 5GC on NAS level are expected to be integrity protected from a certain state onwards. However, it is unclear to which level UEs actually implement this part of the specifications and discard messages that are not protected using at least level NIA1. UEs that reply to unprotected Security Mode commands will still expose their IMEI to a rogue network and thus indirectly disclose the identity of the subscriber. Verification of adherence of a UE to the standard could be achieved by modifying a fully functional standalone SDR implementation of a 5G network that allows to deactivate the integrity protection for selected messages and using test SIM cards under the control of the researcher. \\

For data confidentiality, the activation of encryption of data on radio level and on NAS level is entirely under the control of the network operator. It needs to be verified to which extent operators activate RRC, NAS and user plane encryption. If, in the control plane, an operator only relies on integrity protection, then the IMEI/PEI and the associated 5G-GUTI of the device can still leak and allow tracking of the user even if user plane data is encrypted. Using a fully instrumented test UE that provides access to this level of information would allow to verify the protection level used by operators in the field. \\

On network side, it is unclear to which extent operators actually implement IPSec between all network functions. If an operator relies on the physical security of the network links, then this might allow interception of confidential data (including key material) between the network entities. Without physically forcing access to the operator’s network, the use of IPSec can only be verified by auditing the network operators.
\section*{Acknowledgements}
A special thank you goes to the reviewers for their insightful feedback which helped us enhance this report: \\ \\
\noindent
Alain Paschoud, Kudelski SA \\
Nicolas Mutschler, Kudelski SA \\
Benoît Gerhard, Kudelski SA

\clearpage
\bibliographystyle{unsrt}
\bibliography{bibli}

\end{document}